\documentclass{article}
\usepackage{amssymb}

\usepackage[final]{corl_2025} 
\usepackage{graphicx}
\usepackage{subcaption}
\usepackage{titlesec}
\usepackage{wrapfig}
\usepackage{amsmath}

\titlespacing{\section}{0pt}{4pt}{1pt}
\titlespacing{\subsection}{0pt}{3pt}{1pt}
\titlespacing{\subsubsection}{0pt}{2pt}{0.5pt}

\newcounter{myenum}
{\end{list}}

\newenvironment{flushitemize}{%
\begin{list}{$\bullet$}
   {\setlength{\leftmargin}{15pt}}%
    \setlength{\labelwidth}{20pt}
    \setlength{\itemindent}{0pt}
    \setlength{\labelsep}{0.5em}
 \setlength{\itemsep}{1pt}
 \setlength{\parskip}{0pt}
 \setlength{\parsep}{0pt}}
{\end{list}}

\title{CASH: Capability-Aware Shared Hypernetworks for \\ Flexible Heterogeneous Multi-Robot Coordination}

\author{
    Kevin Fu$^*$, Shalin Anand Jain$^*$, Pierce Howell, Harish Ravichandar \\
    Georgia Institute of Technology, Atlanta, USA
}

\begin{document}
\maketitle
\def\thefootnote{*}\footnotetext{Equal Contribution}\def\thefootnote{\arabic{footnote}}
\def\thefootnote{}\footnotetext{Project Website: \url{https://star-lab.cc.gatech.edu/papers/fu-jain-CASH-CoRL/}}\def\thefootnote{\arabic{footnote}}
\begin{abstract}
Recent advances have enabled heterogeneous multi-robot teams to learn complex and effective coordination skills. 
However, existing neural architectures that support heterogeneous teaming tend to force a trade-off between expressivity and efficiency.
Shared-parameter designs prioritize sample efficiency by enabling a single network to be shared across all or a pre-specified subset of robots (via input augmentations), but tend to limit behavioral diversity.
In contrast, recent designs employ a separate policy for each robot, enabling greater diversity and expressivity at the cost of efficiency and generalization. 
Our key insight is that such tradeoffs can be avoided by viewing these design choices as ends of a broad spectrum.
Inspired by recent work in transfer and meta learning, and building on prior work in multi-robot task allocation, we propose Capability-Aware Shared Hypernetworks (CASH), a \textit{soft weight sharing} architecture that uses hypernetworks to efficiently learn a \textit{flexible} shared policy that dynamically adapts to each robot post training.
By explicitly encoding the impact of robot capabilities (e.g., speed and payload) on collective behavior, CASH enables \textit{zero-shot generalization} to unseen robots or team compositions.
Our experiments involve multiple heterogeneous tasks, three learning paradigms (imitation learning, value-based, and policy-gradient RL), and SOTA multi-robot simulation (JaxMARL) and hardware (Robotarium) platforms.
Across all conditions, we find that CASH generates appropriately-diverse behaviors and consistently outperforms baseline architectures in terms of performance and sample efficiency during both training and zero-shot generalization, all with 60$\%$-80$\%$ fewer learnable parameters. 
\end{abstract}
\keywords{Multi-Robot Learning, Heterogeneous Teams, Parameter Sharing} 

\section{Introduction}

\begin{figure}
    \centering
    \includegraphics[width=0.9\columnwidth]{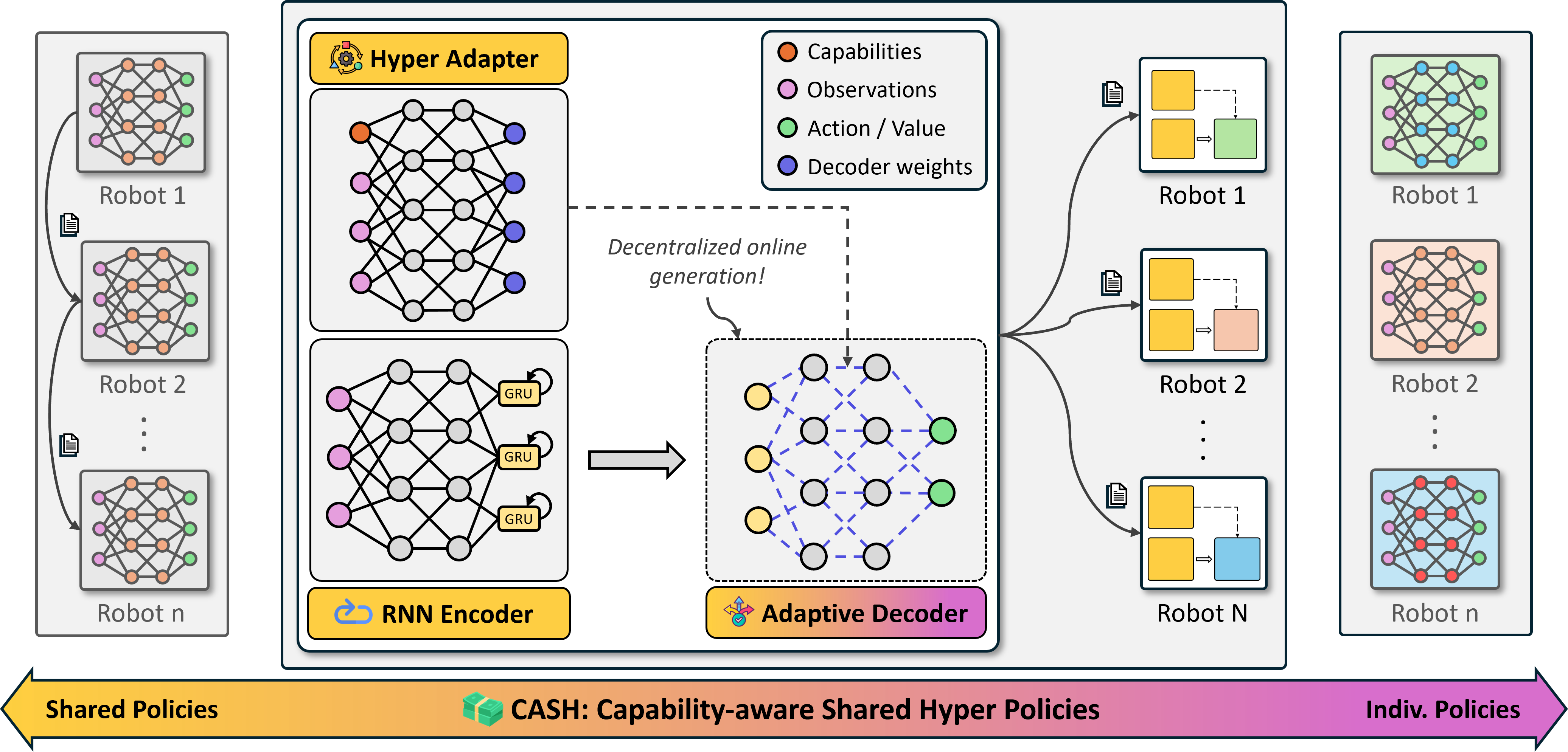}
    \caption{
    We introduce Capability-Aware Shared Hypernetworks (CASH) (\textit{middle}), a novel class of \textit{soft parameter sharing} architectures that establishes and spans the broad spectrum between shared (\textit{left}) and individualized (\textit{right}) parameter designs.
    CASH enables effective decentralized heterogeneous teaming, generalization to unseen robots, diverse behaviors, and greater learning efficiency. 
    }
    \label{fig:CASH-block-diagram}
\end{figure}


Imagine multiple fire departments gathering multiple heterogeneous robots into a team to suppress a wildfire. 
To be effective, the robots must readily adopt diverse roles and reason about how their individual and collective capabilities (e.g., speed, water-capacity, and sensing radius) interact with the environment (e.g., fire intensities and locations). 
Further,  since the team composition is often unknown until runtime and robot capabilities could deteriorate, we need flexible strategies that zero-shot generalize to different team compositions and robot capabilities.

We address the challenge of learning \textit{flexible} and \textit{diverse} coordination strategies that enable \textit{heterogeneous} robots to adapt their behaviors based on their individual and collective \textit{capabilities}. While learning a separate policy for each robot will enable diverse behaviors~\cite{bettiniHeterogeneousMultiRobotReinforcement2023}, they cannot generalize to unseen robots and require significantly more learnable parameters. As such, we aim to encode diverse strategies within a \textit{shared yet flexible} policy, without sacrificing generalization or efficiency.
We propose a novel neural architecture named Capability-Aware Shared Hypernetworks (CASH) to enable flexible and diverse coordination (see Fig.~\ref{fig:CASH-block-diagram}). 
CASH's encoder helps learn \textit{robot-agnostic} coordination strategies that are shared across all robots regardless of their capabilities. Its Hyper Adapter uses a hypernetwork~\cite{Ha_Dai_Le_2016} to determine the weights of its Adaptive Decoder \textit{on-the-fly} based on the current context (i.e, observations) and the robots' capabilities.  Our use of hypernetworks in CASH helps establish \textit{soft weight sharing} architectures within the context of heterogeneous coordination, and encode \textit{robot-} and \textit{context-specific} strategies within a flexible shared architecture. CASH allows some \textit{policy} parameters to vary across robots, while all \textit{learnable} parameters are shared. 

CASH offers several practical benefits. 
First, CASH improves learning efficiency by significantly reducing the number of learnable parameters and sharing them across heterogeneous robots. 
Second, CASH enables zero-shot generalization to robots and team compositions unseen during training.
Third, CASH 
can \textit{automatically} learn appropriate levels of behavioral diversity. 
Fourth, CASH can adapt to \textit{online changes} in robots capabilities. 
Fifth, CASH can be utilized within any learning paradigm (e.g., imitation learning or RL). 
We evaluated CASH on four heterogeneous coordination tasks (wildfire suppression, mining, material transport, and predator-prey) with three separate learning paradigms (imitation learning, value-based RL, and policy-based RL). Our experiments involved a state-of-the-art  MARL framework (JaxMARL~\cite{jaxmarl}) and an established hardware testbed for multi-robot systems (Robotarium~\cite{robotarium}).
Our results demonstrate that CASH consistently outperforms existing  (independent and shared) architectures in terms of sample efficiency, and zero-shot generalization to new robots, new team compositions, and larger teams. Further, we show that these benefits extend to physical deployments, where CASH can adapt to online changes to robot capabilities.
Notably, CASH offers such improvements while using 60$\%$ to 80$\%$ fewer learnable parameters than existing architectures. 

\section{Related Works}\label{sec:related_work}



\textbf{Architectures for Heterogeneous Coordination:} 
Shared-parameter architectures -- where all agents share a single set of learnable parameters -- improve cooperation, learning efficiency, and scalability~\cite{tan1993multi,gupta2017cooperative}.
However, sharing parameters can prohibit diverse agent behaviors and unique roles. While this limitation is often immaterial for homogeneous teams (e.g., formation control~\cite{khanGraphPolicyGradients2020} and path planning~\cite{sartoretti_2019_PRIMALPathfindingReinforcement,li_2020_GraphNeuralNetworks}), it can critically limit heterogeneous teams. 
To extend shared-parameter architectures to heterogeneous teams, prior work has 
simply appended unique information about each agent to the input~\cite{terryRevisitingParameterSharing2023, howell_2023_GeneralizationHeterogeneousMultiRobot};
typically, a \textit{unique ID} assigned to each agent.
While such ID-based designs tend to improve efficiency and scalability, recent work has demonstrated that they often fail to learn sufficient behavioral diversity and are not robust to noisy environments~\cite{christianosScalingMultiAgentReinforcement2021,bettiniHeterogeneousMultiRobotReinforcement2023}.


In contrast, \textit{individualized policies}~\cite{bettiniHeterogeneousMultiRobotReinforcement2023,seraj_2022_LearningEfficientDiversea} (also referred to as independent or heterogeneous policies) dedicate a separate policy for each agent with no parameter sharing
to enable robust and diverse behaviors~\cite{li_2021_CelebratingDiversityShared}. Naturally, this improvement comes at the cost of training efficiency due to significantly more learnable parameters and  limited data reuse.
Further, these approaches do not allow for zero-shot generalization to robots not seen in training.



Prior work has also explored a hybrid approach known as \textit{selective} parameter sharing in which shared parameters are constrained to groups of agents that share a certain characteristic (e.g., robot type~\cite{seraj_2022_LearningEfficientDiversea}, action space~\cite{wakilpoorHeterogeneousMultiAgentReinforcement2020}, inferred role~\cite{christianosScalingMultiAgentReinforcement2021})
with no sharing between groups. 
These approaches are more efficient than individual policy architectures and generate diverse behaviors unlike shared-parameter architectures. However, they too cannot generalize to new robots and assume a small number of known robot types.

We view shared and individualized-parameter architectures as two ends of a broad spectrum with selective sharing falling somewhere in the middle. Within this context, CASH establishes a new class of \textit{soft weight sharing} architectures that can \textit{adaptively} span this spectrum to simultaneously achieve sample efficiency and behavioral diversity while also generalizing to unseen robots.

\textbf{Hypernetworks in multi-agent learning}:
The most notable use of hypernetworks in multi-agent learning is QMIX~\cite{rashid2020monotonic}, a popular off-policy MARL algorithm. QMIX uses hypernetworks to mix individual value estimates to improve decentralized coordination via centralized training. 
In contrast, our novelty is in using hypernetworks to flexibly determine parameters within \textit{individual} robots' policy or value networks based on their capabilities and local observations, enabling a single architecture to encode diverse and adaptive behaviors. 
To the best of our knowledge, there is only one other parallel work that employs hypernetworks within multi-agent policies~\cite{tessera2025hypermarl} which conditions the hypernetwork on agent-specific IDs to produce diverse policies while retaining parameter efficiency. Unlike this work, our design allows for generalization to unseen robots since we condition the hypernetwork on robot capabilities and observations and not on assigned or learned IDs.

\section{Problem Formulation and Objectives}

We model our problem as a Decentralized Partially Observable Markov Decision Process (Dec-POMDP)~\cite{oliehoek2016concise} defined by a tuple $(\mathcal{D}, \mathcal{S}, \mathcal{A}, \mathcal{O}, O, \mathcal{R}, \mathcal{T})$, where $\mathcal{D} = \{1, ..., n\}$ is the set of robots in the team, $\mathcal{S}$ is the set of global states, $\mathcal{A} = \times_{i \in \mathcal{D}} \mathcal{A}_i$ is the joint action space, $\mathcal{O} = \times_{i \in \mathcal{D}} \mathcal{O}_i$ is the joint observation space, $O = P(o | s)$ is the observation function, where $o$ is the joint observation for all robots $\{o_1, ..., o_n  \}$, $\mathcal{R}$ is the reward function, and $\mathcal{T} = P(s' | s,a)$ is the transition function. 

At each timestep $t$, every robot receives an observation $o^t_i$ from the joint observation $o^t \sim O(\cdot|s^t)$. 
Each robot then acts according to their decentralized policy $a^t_i \sim \pi_i(o^t_i)$, forming a joint action $a^t = \{ a_1, ..., a_n\}$. 
The optimal solution to the Dec-POMDP with a fully cooperative team is the set of policies $\{\pi^{*}_1, ..., \pi^{*}_n\}$ that maximize the team's expected reward  $\mathbb{E}[\sum^T_{t=0} r_t]$.

We specifically focus on teams where the robots have heterogeneous capabilities (e.g., speed and sensing radius)~\cite{ravichandar2020strata, howell_2023_GeneralizationHeterogeneousMultiRobot}; such nominal capabilities as we define them can be readily obtained from robot specifications or sensors.
To enable a richer specification of robot heterogeneity, we use 
$C^t = \{c^t_1, ..., c^t_n \}$ to denote the team's capabilities, 
where $c^t_i \in \mathbb{R}^m$
represents the capabilities of $i$th robot at time $t$.
Therefore, we modify the Dec-POMDP definition by giving each policy $\pi_i(\cdot)$ access to the team's collective capabilities, 
yielding $a^t_i \sim \pi_i(o^t_i, C^t)$. 

Finally, our objective is to design a single shared-parameter policy architecture ($\pi = \pi_i, \forall i$) that can produce diverse behaviors and generalize to unseen robots by reasoning about robot capabilities. See Sec. \ref{sec:related_work} for a discussion of existing policy designs.

\section{CASH: Capability-Aware Shared Hypernetworks}
\label{sec:method}

In this section, we introduce CASH, a new class of shared-parameter architectures for flexible and generalizable heterogeneous coordination. 
CASH contains three primary modules (see Fig. \ref{fig:CASH-block-diagram}).

\vspace{3pt}
\noindent \textbf{RNN Encoder}:
In line with contemporary designs~\cite{yuSurprisingEffectivenessPPO2022, rashid2020monotonic}, CASH encodes local observations using a gated recurrent network (GRU)~\cite{cho2014learningphraserepresentationsusing} to handle longer time-horizons and partial observability.
The latent embedding generated by this encoder is passed as input to the Adaptive Decoder.

\vspace{3pt}
\noindent \textbf{Adaptive Decoder}:
Following standard practice, we implement our decoder as either a single layer or a two-layer MLP depending on the learning paradigm (see Appendix \ref{sec:training-details} for details).
 Note that only the \textit{structure} of our decoder is identical across robots. Unlike prior designs, CASH allows the decoder's \textit{parameters} (determined by the Hyper Adapter) to be unique to each robot and context.


\vspace{3pt}
\noindent \textbf{Hyper Adapter}:
To allow the Adaptive Decoder to flexibly adapt to each robot and its context, we generate its weights using a hypernetwork-based Adapter module. To represent the current context, we condition the Hyper Adapter on the ego-robot's capabilities, the team's capabilities, and the ego observation. 
For additional implementation details, see Appendix \ref{sec:layernorm} and \ref{subsec:hyper-adapter}.

By using a hypernetwork and sharing \textit{all learnable parameters} across robots, CASH retains the typical benefits of both full parameter sharing (sample efficiency and generalization to unseen robots) and individualized parameters (diverse and effective heterogeneous coordination). Further, CASH supports \textit{decentralization} as it relies only on local observations and capability information. 

We employ hypernetworks~\cite{Ha_Dai_Le_2016} since they tackle two common limitations of standard neural networks: i) lack of ability to adapt knowledge between different data contexts, and ii) performance drops due to distributional shifts. Our use of hypernetworks is inspired by the fact that they can produce neural network parameters that allow a target network (e.g. robot policy) to dynamically adapt to new contexts (e.g. tasks,  robot capabilities, etc.). 
Hypernetworks have been shown to encode policies for multiple tasks within a single architecture~\cite{huang_2021_ContinualModelBasedReinforcementa,rezaei-shoshtari_2023_HypernetworksZeroshotTransfer, vonOswald_Henning_Grewe_Sacramento_2022}, improve learning efficiency by improving gradient estimation in Q-learning~\cite{sarafian_2021_RecomposingReinforcementLearning}, and achieve better performance with significantly fewer learnable parameters in meta-learning~\cite{Galanti_Wolf_2020, beck_vuorio_xiong_whiteson_2022}.
As our experiments reveal, CASH translates these benefits of hypernetworks to heterogeneous multi-robot coordination (see Sec.~\ref{sec:jaxmarl_experiments} and \ref{sec:robotarium_experiments}).

Formally, at each timestep $t$, given the $i$th robot's observations $o^t_i$, the RNN Encoder $f_{\psi}(\cdot)$ produces a latent embedding
$z^t_i = f_{\psi}(o^t_i)$.
Next, the Hyper Adapter $h_\phi(\cdot)$ generates the parameters $\theta^t_i$ of the $i$th robot's Adaptive Decoder: $\theta^t_i = h_{\phi}(o^t_i, c^t_i, C^t_{/i})$, where $o^t_i$ is the local observation, $c^t_i$ denotes the ego-robot's capabilities, and $C^t_{/i} = \{c^t_j | j \neq i \}$ is teammate capabilities.
Note that both $f_{\psi}(\cdot)$ and $h_\phi(\cdot)$ lack the subscript $i$ as they are parameter-shared across robots. 
Finally, the Adaptive Decoder $g_{\theta^t_i}(\cdot)$ produces final actions (or value estimates) for each robot
$a^t_i = g_{\theta^t_i} (z^t_i)$
based on the latent embedding $z^t_i$.
Though the Hyper Adapter's parameters $\phi$ are shared across robots, the Adaptive Decoder's parameters $\theta^t_i$ are unique to each robot.
This can be seen as \textit{soft parameter sharing}~\cite{Chauhan_Zhou_Lu_Molaei_Clifton_2023} applied to multi-robot teams. 
Moreover, since all $\theta^t_i$ are generated at each timestep, CASH can dynamically adapt to online changes to robot capabilities. 

\section{Experimental Evaluation}

We evaluated and compared CASH against baseline architectures using two established experimental platforms: JaxMARL~\cite{jaxmarl} 
(Sec. \ref{sec:jaxmarl_experiments}) and the Robotarium \cite{robotarium} (Sec. \ref{sec:robotarium_experiments}). 
All plots are smoothed by downsampling the original mean and standard deviation over all seeds, then taking a rolling average.

\subsection{Experiments on JaxMARL}

\label{sec:jaxmarl_experiments}



Here, we investigate how CASH i) compares to individual policy designs, 
ii) compares to other shared-parameter designs,
and iii) generalizes to unseen team compositions and robot capabilities. 


\label{subsec:exp-design-jaxmarl}

\noindent \textbf{Architecture Variants}:
We compared CASH against three baseline architectures that reflect SOTA practices to handle heterogeneity. 
To ensure fairness, all baselines and CASH share a commonly used design: an RNN encoder followed by an MLP decoder~\cite{rashid2020monotonic, yuSurprisingEffectivenessPPO2022} (see Appendix \ref{sec:training-details}).
\vspace{-2pt}
\begin{flushitemize}
    \item INDV: Each robot employs a separate architecture without sharing parameters (e.g., ~\cite{bettiniHeterogeneousMultiRobotReinforcement2023}). This baseline questions if our shared-parameter approach trades performance for efficiency.
    \item RNN-IMP: The standard RNN-based architecture that is \textit{implicitly} conditioned on capabilities by way of processing observations.
    Indeed, some capabilities (e.g. speed) can be inferred solely from observations over time with memory-enabled architectures (e.g., GRU). 
    This baseline questions the need for explicit conditioning on capabilities.
    \item RNN-EXP: A modification of the standard RNN architecture that \textit{explicitly} considers capabilities by appending them to observations. 
    This baseline emulates prior designs that append agent-specific information~\cite{terryRevisitingParameterSharing2023, howell_2023_GeneralizationHeterogeneousMultiRobot} and questions the need for hypernetworks.
    \item CASH: Our proposed architecture also explicitly considers capability information, but leverages a hypernetwork to dynamically adapt the action decoder to the robot and context (Sec.~\ref{sec:method}).
\end{flushitemize} 
\vspace{-2pt}

\begin{figure}
    \centering
    \begin{subfigure}{0.24\columnwidth}
      \includegraphics[width=\columnwidth]{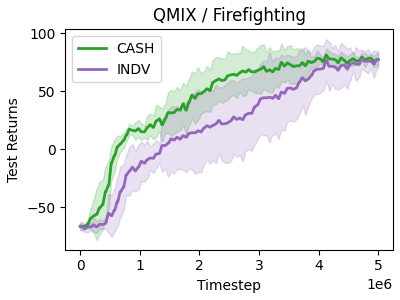}
    \end{subfigure}
    \begin{subfigure}{0.24\columnwidth}
      \includegraphics[width=\columnwidth]{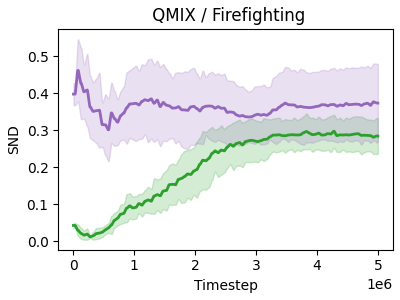}
    \end{subfigure}
    \begin{subfigure}{0.24\columnwidth}
      \includegraphics[width=\columnwidth]{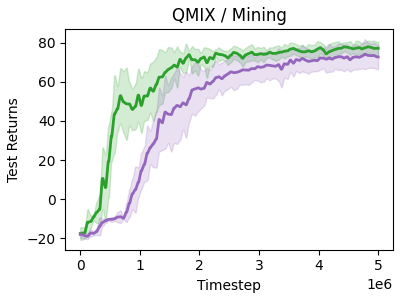}
    \end{subfigure}
    \begin{subfigure}{0.24\columnwidth}
      \includegraphics[width=\columnwidth]{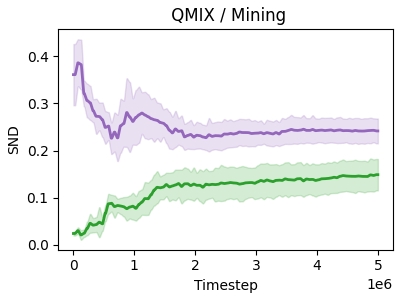}
    \end{subfigure}
    \begin{subfigure}{0.24\columnwidth}
      \includegraphics[width=\columnwidth]{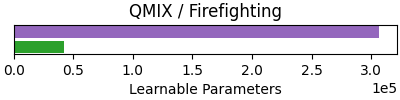}
    \end{subfigure}
    \begin{subfigure}{0.24\columnwidth}
      \includegraphics[width=\columnwidth]{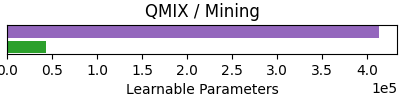}
    \end{subfigure}
    
    \caption{
    CASH is more sample efficient than individualized policies (see returns) and learns more-effective levels of diversity (see SND), while using drastically fewer learnable parameters (bottom).
    }
    \label{fig:ps-ablation}
\end{figure}

\noindent \textbf{Learning Paradigms}:
We train CASH with three standard approaches for multi-agent learning, QMIX, MAPPO, and DAgger (representing off-policy MARL, on-policy MARL, and imitation learning, respectively), to demonstrate that CASH is effective regardless of training algorithm. 
During training, we assume centralized access to the joint observations and actions of all robots, though each robot policy is decentralized (i.e. operates only on local information $o^t_i$ and easily updated capability information $C^t$).
This is a common multi-robot learning assumption known as centralized training, decentralized execution (CTDE) 
~\cite{kraemer_2016_MultiagentReinforcementLearning, loweMultiAgentActorCriticMixed2017}.

For MAPPO and QMIX, we adapt JaxMARL's~\cite{jaxmarl} existing implementations to include capabilities. We contribute an implementation of DAgger for JaxMARL based on the existing MARL implementations.
All results involving QMIX and MAPPO are over 10 random seeds; all DAgger results are over 3 random seeds.
For specific training/architecture details, see Appendix \ref{sec:training-details} and our code implementation.




\noindent \textbf{JaxMARL Tasks}:
We evaluate our method on two heterogeneous cooperative tasks. 
Both environments are implemented with JaxMARL's variant of the Multi-Agent Particle Environment ~\cite{jaxmarl, loweMultiAgentActorCriticMixed2017}.
For additional experiments on larger team sizes, please see Appendix \ref{sec:jaxmarl-12-agent}.

\texttt{Firefighting}:
A team of three robots spawns in a central depot and aim to put out two fires. 
Robots have varying speed and water capacity, and fires spawn in random positions with varying sizes.
Task success is defined as both fires being successfully extinguished within the time limit.

\texttt{Mining}:
A team of four robots must mine resources from two deposit zones and bring them to a dropoff zone until a quota for each resource is reached.
Robots vary in their carrying capacity for each resource. 
The team succeeds if both quotas are met within a time limit. 
This task is similar to the HMT task in a prior study on capability-awareness in multi-agent teams \cite{howell_2023_GeneralizationHeterogeneousMultiRobot}.



\noindent \textbf{Metrics}: i) \textit{Training returns} ($\uparrow$) is the sum of all rewards collected during training episodes, ii) \textit{Success rate} ($\uparrow$), where success is defined for each task as above, and iii) \textit{SND} ($\updownarrow$) quantifies the behavioral heterogeneity of independent policies~\cite{snd}. We slightly modify SND to accommodate shared-parameter policies conditioned on capabilities (see Appendix \ref{sec:SND}).

\noindent \textbf{Training and Testing Teams}:
We standardize the teams sampled during training and evaluation across CASH and all baselines as follows (see Appendix~\ref{sec:cap-sample} for more details):

\textit{Training teams (in-distribution)}: We sample a fixed number of training teams from a pool of robots whose capabilities uniformly cover a set range. 

\textit{Unseen teams (out-of-distribution)}: We generate unseen team compositions with out-of-distribution robots by sampling some capabilities from ranges outside the training ranges.

Below, we organize our discussion into key observations and experiments.

\textbf{CASH improves sample efficiency and enables generalization without sacrificing performance.}
\label{subsec:cash-vs-indp}
To investigate the impact of soft parameter sharing on performance and behavioral diversity, we compared CASH (soft parameter sharing) against the INDV (individualized parameters) baseline.
Since INDV assumes that the robot team doesn't change, we trained each seed on a \textit{single} unique multi-robot team throughout training.

We find that CASH is more sample-efficient and performs marginally better than INDV in both tasks, as seen by the return curves in Fig. \ref{fig:ps-ablation},
despite having a fraction of the learnable parameters.
The SND plots in Fig. \ref{fig:ps-ablation} show CASH has lower behavioral diversity than INDV.
However, since this lower diversity doesn't result in worse task performance, we conclude that CASH is learning the \textit{appropriate} level of behavioral diversity while INDV produces \textit{superfluous} behavioral diversity that doesn't improve performance.
Critically, unlike INDV, CASH can generalize to unseen robots.

In the experiments reported below, we compare CASH's ability to generalize against the two shared-parameter baselines (RNN-IMP and RNN-EXP) on two heterogeneous coordination tasks across three learning paradigms. Note that we cannot compare against INDV since independent parameter designs do not generalize to new robots. 

\begin{figure*}[b]
    \centering
    \begin{subfigure}{0.24\columnwidth}
      \includegraphics[width=\columnwidth]{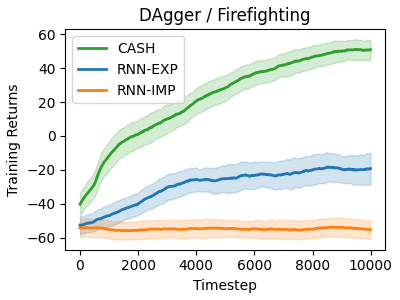}
    \end{subfigure}%
    \begin{subfigure}{0.24\columnwidth}
      \includegraphics[width=\columnwidth]{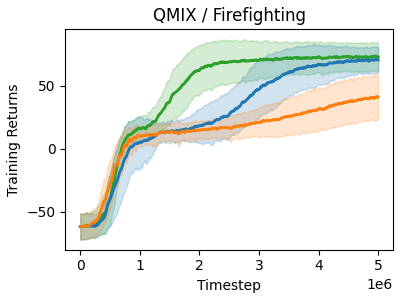}
    \end{subfigure}%
    \begin{subfigure}{0.24\columnwidth}
      \includegraphics[width=\columnwidth]{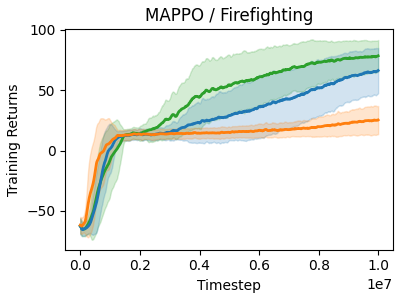}
    \end{subfigure}%
    \begin{subfigure}{0.24\columnwidth}
        \begin{subfigure}{\columnwidth}
          \includegraphics[width=\columnwidth]{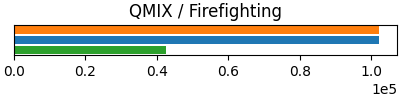}
        \end{subfigure}
        \begin{subfigure}{\columnwidth}
          \includegraphics[width=\columnwidth]{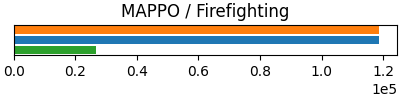}
        \end{subfigure}
        \begin{subfigure}{\columnwidth}
          \includegraphics[width=\columnwidth]{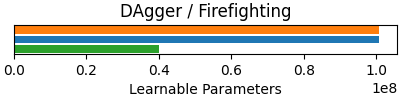}
        \end{subfigure}
    \end{subfigure}

    \begin{subfigure}{0.24\columnwidth}
      \includegraphics[width=\columnwidth]{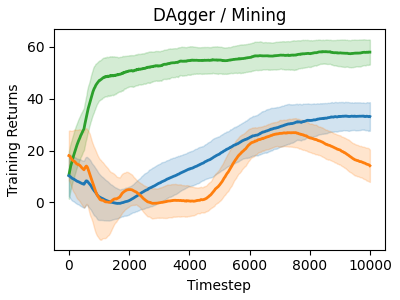}
    \end{subfigure}%
    \begin{subfigure}{0.24\columnwidth}
      \includegraphics[width=\columnwidth]{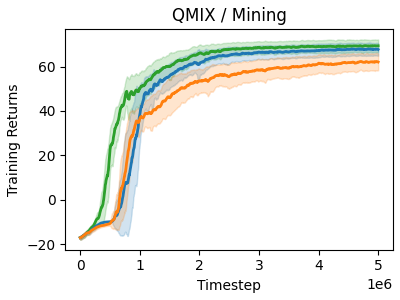}
    \end{subfigure}%
    \begin{subfigure}{0.24\columnwidth}
      \includegraphics[width=\columnwidth]{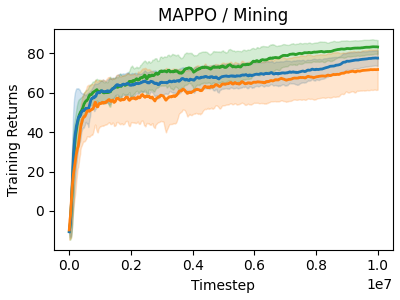}
    \end{subfigure}%
    \begin{subfigure}{0.24\columnwidth}
        \begin{subfigure}{\columnwidth}
          \includegraphics[width=\columnwidth]{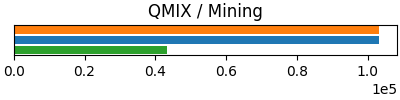}
        \end{subfigure}
        \begin{subfigure}{\columnwidth}
          \includegraphics[width=\columnwidth]{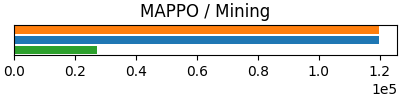}
        \end{subfigure}
        \begin{subfigure}{\columnwidth}
          \includegraphics[width=\columnwidth]{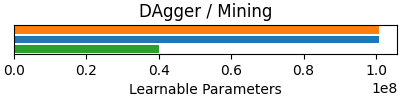}
        \end{subfigure}
    \end{subfigure}
    
    \caption{
    Across two tasks and three learning paradigms, CASH is consistently more sample efficient and yields better returns than the baselines despite using 60\%-80\% fewer learnable parameters. 
    }
    \label{fig:train-return-and-param-ct}
\end{figure*}

\textbf{CASH improves parameter and sample efficiency for shared-parameter methods.}
\label{jaxmarl-efficiency}
As seen in Fig~\ref{fig:train-return-and-param-ct}, CASH is consistently the most sample-efficient method and achieves the highest returns while using 60-80\% fewer parameters.
CASH's improved efficiency can be attributed to its hypernetwork better modeling the influence of robot capabilities on the team's decisions.
Note that the relative benefits of CASH are exaggerated for imitation learning, in which each architecture receives two to three orders of magnitude fewer samples than in the RL settings. This suggests that CASH can better handle data-scarce regimes compared to baselines. 
These trends hold for task-specific metrics as well (see Appendix \ref{sec:additional-jaxmarl-results}).


\begin{table*}[]
    \centering
    \resizebox{\textwidth}{!}{%
    \begin{tabular}{|c|c|ccc|ccc|}
        \hline 
        & & & \texttt{Firefighting} & & &\texttt{Mining} & \\
         Algorithm & Architecture & In-Distribution & Out-of-Distribution & Out-of-Distribution & In-Distribution & Out-of-Distribution & Out-of-Distribution \\
          & & Success Rate ($\uparrow$) & Success Rate ($\uparrow$) & SND & Success Rate ($\uparrow$) & Success Rate ($\uparrow$) & SND \\
        \hline
         QMIX & RNN-IMP & 0.54 $\pm$ 0.28 & 0.49 $\pm$ 0.22 & 0.00 $\pm$ 0.00 & 0.79 $\pm$ 0.08 & 0.63 $\pm$ 0.12 & 0.00 $\pm$ 0.00 \\
            & RNN-EXP & \textbf{0.97 $\pm$ 0.05} & 0.56 $\pm$ 0.26 & 0.15 $\pm$ 0.07 & 0.98 $\pm$ 0.03 & 0.83 $\pm$ 0.15 & 0.07 $\pm$ 0.00 \\
            & CASH & 0.96 $\pm$ 0.10 & \textbf{0.67 $\pm$ 0.09} & 0.28 $\pm$ 0.03 & \textbf{1.00 $\pm$ 0.00} & \textbf{0.88 $\pm$ 0.12} & 0.16 $\pm$ 0.01 \\
        \hline
         MAPPO & RNN-IMP & 0.21 $\pm$ 0.20 & 0.15 $\pm$ 0.14 & 0.00 $\pm$ 0.00 & 0.61 $\pm$ 0.24 & 0.43 $\pm$ 0.17 & 0.00 $\pm$ 0.00 \\
            & RNN-EXP & \textbf{0.71 $\pm$ 0.30} & 0.43 $\pm$ 0.27 & 0.64 $\pm$ 0.03 & \textbf{1.00 $\pm$ 0.00} & 0.74 $\pm$ 0.11 & 0.18 $\pm$ 0.03 \\
            & CASH & \textbf{0.71 $\pm$ 0.43} & \textbf{0.68 $\pm$ 0.17} & 0.58 $\pm$ 0.03 & 0.98 $\pm$ 0.04 & \textbf{0.83 $\pm$ 0.12} & 0.22 $\pm$ 0.06 \\
        \hline
         DAgger & RNN-IMP & 0.08 $\pm$ 0.14 & 0.00 $\pm$ 0.00 & 0.00 $\pm$ 0.00 & 0.00 $\pm$ 0.00 & 0.07 $\pm$ 0.08 & 0.00 $\pm$ 0.00 \\
            & RNN-EXP & 0.17 $\pm$ 0.14 & 0.08 $\pm$ 0.14 & 0.02 $\pm$ 0.00 & 0.35 $\pm$ 0.10 & 0.30 $\pm$ 0.10 & 0.03 $\pm$ 0.00 \\
            & CASH & \textbf{1.00 $\pm$ 0.00} & \textbf{0.92 $\pm$ 0.14} & 0.08 $\pm$ 0.01 & \textbf{0.88 $\pm$ 0.11} & \textbf{0.83 $\pm$ 0.10} & 0.06 $\pm$ 0.00 \\
        \hline
    \end{tabular}}
    \caption{
    CASH tends to achieve the highest success rates across all JaxMARL tasks and learning paradigms, particularly when evaluated on teams with out-of-distribution robot capabilities.
    }
    \label{tab:success_rates}
\end{table*}

\textbf{CASH improves generalization to unseen team compositions and capabilities.}
\label{jaxmarl-generalization}
We next evaluated CASH's capacity for zero-shot generalization to unseen team compositions and capabilities, again comparing against the two shared-parameter baselines (RNN-IMP and RNN-EXP) across all conditions (as INDV cannot generalize to new robots). 
We report success rates and SND across conditions in Table \ref{tab:success_rates}.
Out-of-distribution success rates are from evaluation on unseen test teams (see Sec.~\ref{subsec:exp-design-jaxmarl}), while in-distribution success rates are from training teams.

While all architectures perform worse when generalizing to new teams (as expected), CASH exhibits the lowest drop in performance as a result of its superior capability grounding.
Notably, when trained with QMIX or MAPPO, RNN-EXP and CASH result in similar success rates on training teams.
However, within the same task and learning paradigm, CASH vastly outperforms RNN-EXP on unseen teams.
These findings show that the baselines can learn effective strategies to cope with heterogeneity between agents in training teams, but struggle to generalize them to unseen teams.
By contrast, CASH learns generalizable strategies that exploit the relationship between robot capability and desired behavior associated with each task.

We also note that CASH tends to produce greater behavioral diversity as measured by SND. 
This increased diversity partially likely explains CASH's superior generalization to unseen capabilities -- effective generalization may require policies to generate different behaviors when robot capabilities change, even if observations remain the same.
Such differentiation would not be possible for a policy with very limited behavioral diversity. 
We hypothesize that behavioral diversity might be a necessary but not sufficient condition for successful generalization since greater diversity doesn't always correlate with better performance (e.g., see comparison against INDV. in Sec.~\ref{subsec:cash-vs-indp}).

\subsection{Experiments on the Robotarium}
\label{sec:robotarium_experiments}
To validate the applicability of CASH to real multi-robot systems, we evaluate it on the Robotarium \cite{robotarium}, a publicly available multi-robot physical testbed and simulator that provides realistic robot dynamics and barrier certificates. 
We again evaluate CASH against the two shared-parameter baselines (RNN-IMP and RNN-EXP).
We train all baselines with QMIX using the MARBLER platform \cite{torbati2023marbler}, which bridges the Robotarium's simulator with existing MARL implementations in EPyMARL \cite{papoudakis2021benchmarking} and provides several multi-robot tasks to evaluate policies (see Appendix \ref{sec:training-details} for training details). 

\begin{wrapfigure}[17]{r}{0.35\columnwidth}
    \centering
    \begin{subfigure}{0.35\columnwidth}
      \includegraphics[width=\columnwidth]{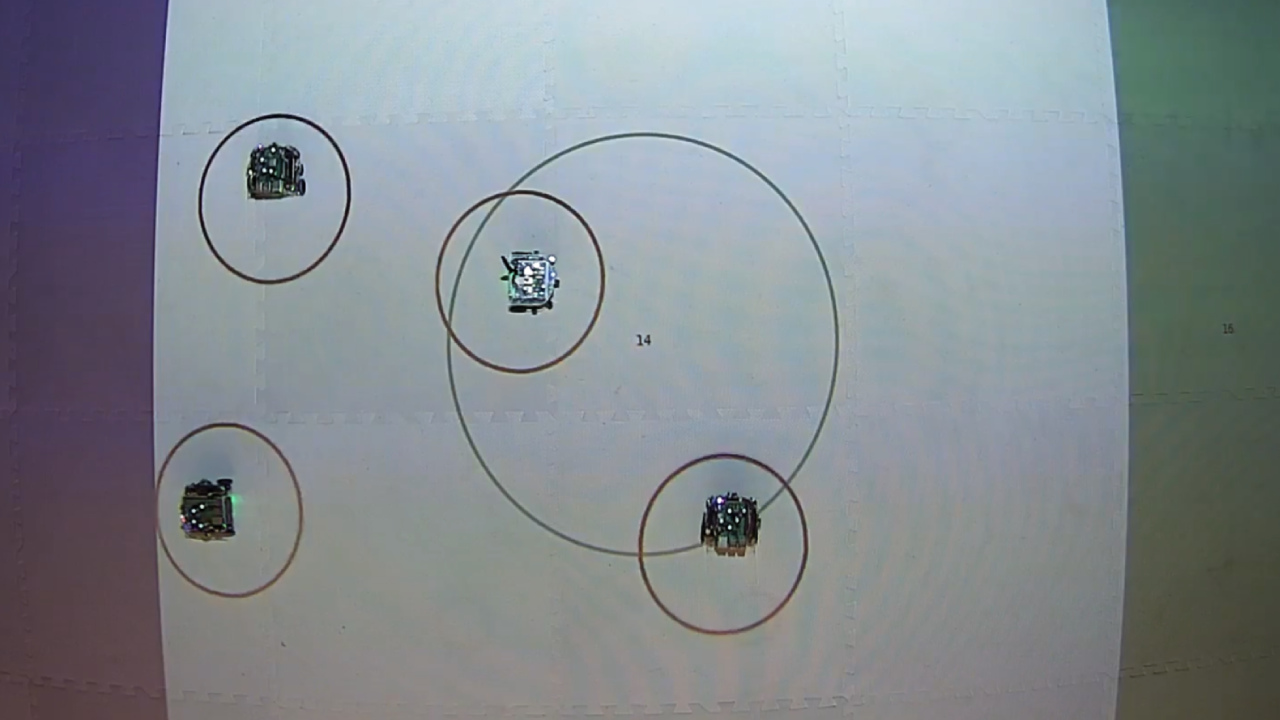}
    \end{subfigure}

    \begin{subfigure}{0.35\columnwidth}
      \includegraphics[width=\columnwidth]{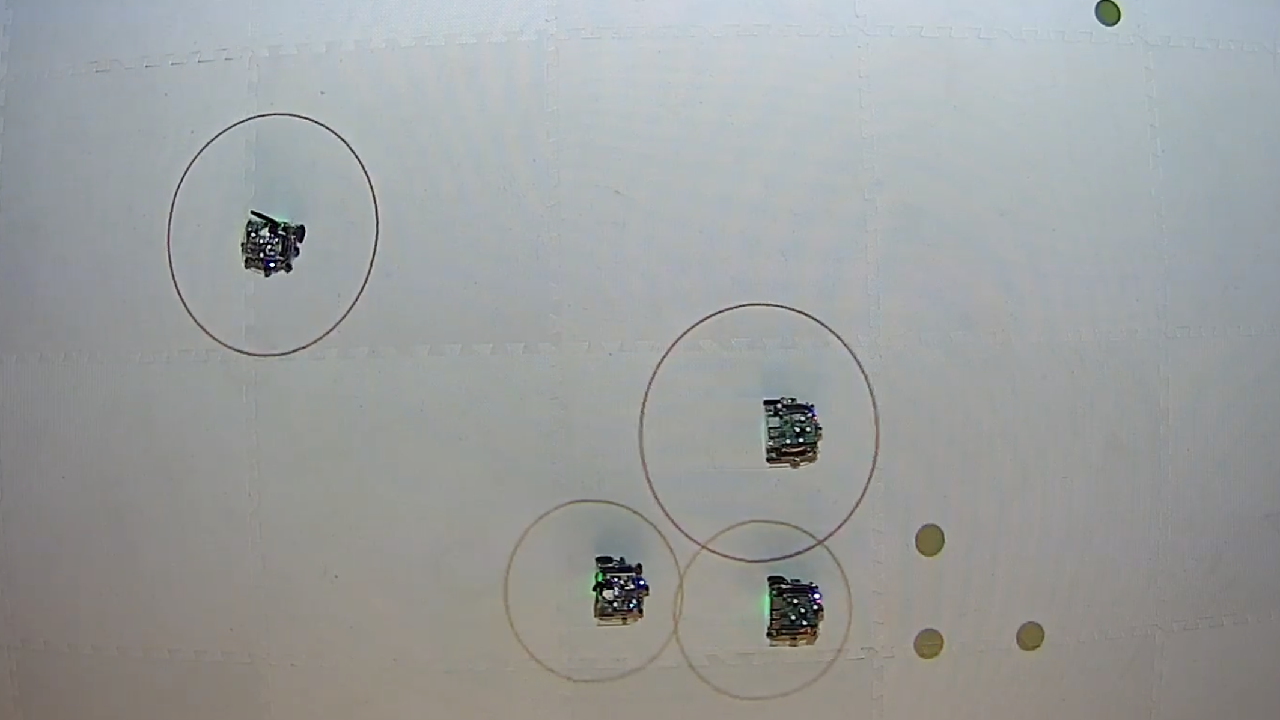}
    \end{subfigure}
    \caption{
    Snapshots of physical deployment on two Robotarium tasks: \texttt{MT} (top), \texttt{PCP} (bottom).
    }
    \label{fig:sim-to-real}
\end{wrapfigure}

\vspace{3pt}
\noindent \textbf{Tasks}:
We evaluate on two established heterogeneous multi-robot tasks from MARBLER~\cite{torbati2023marbler}.
For additional experiments on these tasks with larger team sizes, see Appendix \ref{sec:marbler-12-agent}.

\texttt{Material Transport (MT)}: A team of four robots must unload materials from two zones (green circle and rectangle) into a dropoff zone (purple). 
The team is rewarded for amount of material loaded and unloaded, penalized for collisions, and penalized each timestep the loading zones have remaining material. Robots vary in their speed and carrying capacity.

\texttt{Predator Capture Prey (PCP)}: Two sensing robots and two capture robots must collaborate to capture prey (circular markers).
Sensing and capturing prey is rewarded while collisions and time-to-capture are penalized.
Robots vary in their sensing and capture radii.

\textbf{Online Adaptation:} Additionally, to investigate how CASH handles \textit{online} changes to robot capabilities, we create two special scenarios within the \texttt{MT} and \texttt{PCP} tasks: i) \texttt{Failure}, where halfway through the episode, a robot is selected randomly and its capability (speed in \texttt{MT}, sensing/capture radius in \texttt{PCP}) is decreased by $75\%$ and ii) \texttt{Battery Drain}, where at each step, the capabilities of all robots (speed in \texttt{MT}, sensing/capture radius in \texttt{PCP}) are decayed by a discount factor. Exact capability ranges and deployment teams for these tasks are detailed in Appendix \ref{sec:cap-sample}.


\vspace{3pt}

\noindent \textbf{Metrics}: We report the following metrics for all tasks, averaged across evaluation episodes: i) \textit{Reward} ($\uparrow$), the sum of rewards collected per episode, ii) \textit{Makespan} ($\downarrow$), the episode length, and iii) \textit{Collisions} ($\downarrow$), the number of collisions per episode. Simulation metrics are averaged across 3 seeds, 500 episodes per seed. The highest performing model weights for each baseline are selected for physical robot deployment and metrics are averaged across 5 episodes.

\begin{table*}[]
\centering
\resizebox{\textwidth}{!}{%
\begin{tabular}{|c|c|c|ccc|ccc|}
\hline
& & & &\textit{Robotarium} & & & \textit{Robotarium} & \\
& & & &\textit{(sim - 3 seeds)} & & & \textit{(real - best seed)} & \\
\hline
Task &Architecture &\# Parameters ($\downarrow$) &Reward ($\uparrow$) &Makespan ($\downarrow$) &Collisions ($\downarrow$) &Reward ($\uparrow$) &Makespan ($\downarrow$) &Collisions ($\downarrow$) \\\hline
\texttt{MT} &RNN-IMP &401K &2.32 $\pm$ 11.86 &63.85 $\pm$ 12.41 &0.03 $\pm$ 0.17 &21.52 $\pm$ 1.57 &54.20 $\pm$ 3.92 &\textbf{0.00 $\pm$ 0.00} \\
&RNN-EXP &401K &12.94 $\pm$ 8.21 &53.59 $\pm$ 13.10 &0.02 $\pm$ 0.13 &23.12 $\pm$ 2.96 &50.20 $\pm$ 7.39 &\textbf{0.00 $\pm$ 0.00} \\
&CASH &\textbf{162K} &\textbf{14.84 $\pm$ 7.91} &\textbf{49.79 $\pm$ 12.87} &\textbf{0.01 $\pm$ 0.12} &\textbf{23.84 $\pm$ 3.72} &\textbf{48.40 $\pm$ 9.31} &\textbf{0.00 $\pm$ 0.00} \\\hline

\texttt{PCP} &RNN-IMP &402K &64.65 $\pm$ 31.923 &79.72 $\pm$ 6.29  &0.03 $\pm$ 0.18 &69.40 $\pm$ 32.65 &\textbf{81.00 $\pm$ 0.00} &\textbf{0.00 $\pm$ 0.00} \\
&RNN-EXP &402K &61.05 $\pm$ 30.54 &79.60 $\pm$ 6.75 &0.03 $\pm$ 0.17 &31.00 $\pm$ 33.12 &\textbf{81.00 $\pm$ 0.00} &\textbf{0.00 $\pm$ 0.00} \\
&CASH &\textbf{164K} &\textbf{89.21 $\pm$ 33.54} &\textbf{76.86 $\pm$ 9.28} &\textbf{0.02 $\pm$ 0.13} &\textbf{96.60 $\pm$ 18.49} &\textbf{81.00 $\pm$ 0.00} &\textbf{0.00 $\pm$ 0.00} \\
\hline
\end{tabular}}
\caption{
CASH achieves the highest reward, lowest makespan, and fewest collisions on hardware.
}
\label{tab:sim2real}
\end{table*}

\begin{table*}[t]
\centering
\resizebox{\textwidth}{!}{%
\begin{tabular}{|c|c|ccc|ccc|}
\hline
& & &\texttt{MT} & & &\texttt{PCP} & \\
Dynamic Change & Architecture &Reward ($\uparrow$) &Makespan ($\downarrow$) &Collisions ($\downarrow$) &Reward ($\uparrow$) &Makespan ($\downarrow$) &Collisions ($\downarrow$) \\
\hline
\texttt{Battery Drain} &RNN-IMP &-0.39 $\pm$ 11.98 &66.79 $\pm$ 9.43 &0.03 $\pm$ 0.18 &61.09 $\pm$ 31.51 &79.93 $\pm$ 5.97 &0.03 $\pm$ 0.18 \\
&RNN-EXP &9.18 $\pm$ 8.55 &60.01 $\pm$ 12.42 &0.02 $\pm$ 0.13 &57.02 $\pm$ 29.90 &79.82 $\pm$ 6.53 &\textbf{0.03 $\pm$ 0.17}\\
&CASH &\textbf{11.26 $\pm$ 7.91} &\textbf{56.87 $\pm$ 13.29} &\textbf{0.01 $\pm$ 0.12} &\textbf{82.56 $\pm$ 34.25} &\textbf{77.75 $\pm$ 8.85} &\textbf{0.03 $\pm$ 0.17} \\
\hline
\texttt{Failure} &RNN-IMP &0.11 $\pm$ 11.72 &65.73 $\pm$ 11.16 &0.03 $\pm$ 0.17 &56.28 $\pm$ 30.53 &79.95 $\pm$ 6.02 &0.03 $\pm$ 0.18 \\
&RNN-EXP &9.84 $\pm$ 9.22 &57.77 $\pm$ 13.69 &\textbf{0.02 $\pm$ 0.12} &51.10 $\pm$ 29.45 &79.95 $\pm$ 6.56 &0.03 $\pm$ 0.16 \\
&CASH &\textbf{11.74 $\pm$ 8.88} &\textbf{54.77 $\pm$ 14.34} &\textbf{0.02 $\pm$ 0.12} &\textbf{69.89 $\pm$ 34.21} &\textbf{78.95 $\pm$ 7.66} &\textbf{0.02 $\pm$ 0.15} \\
\hline
\end{tabular}}
\caption{
CASH outperforms baselines when adapting to online changes in capabilities.
}
\label{tab:online-cap}
\end{table*}


\textbf{CASH's benefits extend to real multi-robot teams.}
\label{marbler-sim2real}
We find that CASH continues to outperform baselines in training despite having significantly fewer parameters, as seen in Tab. \ref{tab:sim2real}. 
In the \texttt{MT} scenario, we observed that CASH is better at reasoning about the speeds and capacities of robots, resulting in smoother and more direct trajectories between the material loading zones and dropoff zone.
Meanwhile, in \texttt{PCP}, CASH more consistently learns effective role assignment strategies than baselines.
See our supplemental video for full rollouts and more detailed analysis.

For each architecture and task, we selected the highest-performing model across training seeds and physically deployed it on the Robotarium (Fig.~\ref{fig:sim-to-real}). 
We find that CASH effectively coordinates the physical robots and outperforms even the best-performing seeds of the baseline architectures (Tab.~\ref{tab:sim2real}). 
The rewards are typically higher when physically deployed than in simulation since the Robotarium's use of barrier certificates prevents collisions that would have otherwise occurred. 


\textbf{CASH adapts to online changes to robot capabilities.}
\label{subsec:marbler-online-cap}
We investigated how CASH handles online changes to robot capabilities by taking the trained policies for each task and zero-shot deploying them in the $\texttt{Failure}$ and $\texttt{Battery Drain}$ scenarios. 
As expected, we observe a drop in task performance across all baselines in Tab. \ref{tab:online-cap}.
However, CASH consistently outperforms the baselines when dealing with the challenging unplanned changes to robot capabilities and observations during policy execution. 
This is likely due to CASH's better capability grounding that allows it to dynamically adjust the robots' behaviors even during a rollout, a critical feature to enable robustness in learned heterogeneous teaming.



Qualitatively, we notice that CASH better adapts to the changes in robot speed in \texttt{MT} with robots still being effectively allocated and continuing to pick-up and unload material. In contrast, the inefficient navigation strategies of RNN-IMP and RNN-EXP are heavily impacted by the degraded speed, further hindering their performance. Similarly, in the \texttt{PCP} scenarios, CASH's role-assignment strategy better generalizes to the diminished sensing and capturing capability, while the lack of role assignment in RNN-IMP and RNN-EXP prevents the robots from efficiently localizing prey.

\section{Conclusion} 
\label{sec:conclusion}

We proposed a new architecture named \textit{Capability-Aware Shared Hypernetworks (CASH)} that introduces soft parameter sharing in heterogeneous multi-robot learning, establishing a new middle ground between shared and individualized parameter approaches.
CASH is deployable on decentralized teams and supports imitation, value-based, and policy-based RL.


As our experiments on both simulated and real multi-robot platforms reveal, CASH matches the performance of independent-parameter designs while enabling zero-shot generalization and improving sample efficiency. Furthermore, CASH outperforms existing shared-parameter designs in terms of diversity, sample- and parameter-efficiency, zero-shot generalization, and online adaptation.


	



	

\clearpage

\section{Limitations}
\label{sec:limitations}
Though our work represents an exciting step towards flexible and generalizable heterogeneous multi-robot coordination using parsimonious policy architectures, it has some limitations.

One limitation is that our experiments assume lossless communication between neighboring robots.
We simulate true partial observability by directly appending the three nearest teammates' relative positions to the observations of each robot.
While reasonable in smaller-scale controlled settings, this assumption might limit the applicability of CASH to larger-scale multi-robot systems.
Future work could relax this assumption by integrating lossy communication modules into the training process, e.g. by incorporating a GNN~\cite{howell_2023_GeneralizationHeterogeneousMultiRobot, bettiniHeterogeneousMultiRobotReinforcement2023}.

A constraint of CASH is that it cannot handle entirely new capabilities dimensions
during inference. This is a natural limitation of most neural network learning paradigms with the exception of large reasoning models that leverage massive amounts of data and tokenization, which could potentially handle new capabilities during inference time. Furthermore, it is fair to assume for most single-domain policies that the capability dimensions are the same for training and inference.

Another limitation is that the JaxMARL tasks which we evaluated CASH on are simpler than the most complex tasks in heterogeneous multi-agent coordination.
For instance, our \texttt{Firefighting} task can be seen as a simplified version of a firefighting simulator like FireCommander~\cite{seraj_2022_LearningEfficientDiversea}.
However, our results show that despite the simplicity of these tasks, they have the crucial property that better grounding of heterogeneous capabilities tends to yield better performance, as evidenced by RNN-EXP outperforming RNN-IMP and CASH outperforming RNN-EXP.
In our opinion, this is the most important characteristic relevant to heterogeneous robots, and abstracting away other details (e.g. more realistic communication, continuous action spaces, sensing and actuation noise, more dynamic environments) allowed us to tease out how policies can more effectively capture the relationship between robot capabilities and behaviors.
Further, our experiments in the Robotarium show that the benefits of CASH observed in simulation hold for real multi-robot teams as well.

\acknowledgments{
The authors would like to thank the reviewers for their detailed and constructive feedback to help strengthen the paper. 
We would also like to thank the Robotarium by Georgia Tech for it's accessibility to multi-robot hardware for real-robot experimentation and testing.
This work was supported in part by the Army Research Lab under Grant W911NF17-2-0181 (DCIST CRA).
}


\bibliography{camera-ready}  

\begin{thebibliography}{40}
\providecommand{\natexlab}[1]{#1}
\providecommand{\url}[1]{\texttt{#1}}
\expandafter\ifx\csname urlstyle\endcsname\relax
  \providecommand{\doi}[1]{doi: #1}\else
  \providecommand{\doi}{doi: \begingroup \urlstyle{rm}\Url}\fi

\bibitem[Bettini et~al.(2023)Bettini, Shankar, and Prorok]{bettiniHeterogeneousMultiRobotReinforcement2023}
M.~Bettini, A.~Shankar, and A.~Prorok.
\newblock Heterogeneous multi-robot reinforcement learning.
\newblock In \emph{Proceedings of the 2023 International Conference on Autonomous Agents and Multiagent Systems}, 2023.

\bibitem[Ha et~al.(2017)Ha, Dai, and Le]{Ha_Dai_Le_2016}
D.~Ha, A.~M. Dai, and Q.~V. Le.
\newblock Hypernetworks.
\newblock In \emph{International Conference on Learning Representations}, 2017.

\bibitem[Rutherford et~al.(2024)Rutherford, Ellis, Gallici, Cook, Lupu, Ingvarsson~Juto, Willi, Hammond, Khan, Schroeder~de Witt, et~al.]{jaxmarl}
A.~Rutherford, B.~Ellis, M.~Gallici, J.~Cook, A.~Lupu, G.~Ingvarsson~Juto, T.~Willi, R.~Hammond, A.~Khan, C.~Schroeder~de Witt, et~al.
\newblock Jaxmarl: Multi-agent rl environments and algorithms in jax.
\newblock \emph{Advances in Neural Information Processing Systems}, 2024.

\bibitem[Wilson et~al.(2020)Wilson, Glotfelter, Wang, Mayya, Notomista, Mote, and Egerstedt]{robotarium}
S.~Wilson, P.~Glotfelter, L.~Wang, S.~Mayya, G.~Notomista, M.~Mote, and M.~Egerstedt.
\newblock The robotarium: Globally impactful opportunities, challenges, and lessons learned in remote-access, distributed control of multirobot systems.
\newblock \emph{IEEE Control Systems Magazine}, 2020.

\bibitem[Tan(1993)]{tan1993multi}
M.~Tan.
\newblock Multi-agent reinforcement learning: Independent vs. cooperative agents.
\newblock In \emph{Proceedings of the Tenth International Conference on Machine Learning}, 1993.

\bibitem[Gupta et~al.(2017)Gupta, Egorov, and Kochenderfer]{gupta2017cooperative}
J.~K. Gupta, M.~Egorov, and M.~Kochenderfer.
\newblock Cooperative multi-agent control using deep reinforcement learning.
\newblock In \emph{Autonomous Agents and Multiagent Systems}, 2017.

\bibitem[Khan et~al.(2020)Khan, Tolstaya, Ribeiro, and Kumar]{khanGraphPolicyGradients2020}
A.~Khan, E.~Tolstaya, A.~Ribeiro, and V.~Kumar.
\newblock Graph policy gradients for large scale robot control.
\newblock In \emph{Proceedings of the Conference on Robot Learning}, 2020.

\bibitem[Sartoretti et~al.(2019)Sartoretti, Kerr, Shi, Wagner, Kumar, Koenig, and Choset]{sartoretti_2019_PRIMALPathfindingReinforcement}
G.~Sartoretti, J.~Kerr, Y.~Shi, G.~Wagner, T.~K.~S. Kumar, S.~Koenig, and H.~Choset.
\newblock Primal: Pathfinding via reinforcement and imitation multi-agent learning.
\newblock \emph{IEEE Robotics and Automation Letters}, 2019.

\bibitem[Li et~al.(2020)Li, Gama, Ribeiro, and Prorok]{li_2020_GraphNeuralNetworks}
Q.~Li, F.~Gama, A.~Ribeiro, and A.~Prorok.
\newblock Graph neural networks for decentralized multi-robot path planning.
\newblock In \emph{2020 IEEE/RSJ International Conference on Intelligent Robots and Systems (IROS)}, 2020.

\bibitem[Terry et~al.(2020)Terry, Grammel, Son, Black, and Agrawal]{terryRevisitingParameterSharing2023}
J.~K. Terry, N.~Grammel, S.~Son, B.~Black, and A.~Agrawal.
\newblock Revisiting parameter sharing in multi-agent deep reinforcement learning.
\newblock \emph{arXiv preprint arXiv:2005.13625}, 2020.

\bibitem[Howell et~al.(2023)Howell, Rudolph, Torbati, Fu, and Ravichandar]{howell_2023_GeneralizationHeterogeneousMultiRobot}
P.~Howell, M.~Rudolph, R.~J. Torbati, K.~Fu, and H.~Ravichandar.
\newblock Generalization of heterogeneous multi-robot policies via awareness and communication of capabilities.
\newblock In \emph{Proceedings of The 7th Conference on Robot Learning}, 2023.

\bibitem[Christianos et~al.(2021)Christianos, Papoudakis, Rahman, and Albrecht]{christianosScalingMultiAgentReinforcement2021}
F.~Christianos, G.~Papoudakis, M.~A. Rahman, and S.~V. Albrecht.
\newblock Scaling multi-agent reinforcement learning with selective parameter sharing.
\newblock In \emph{Proceedings of the 38th International Conference on Machine Learning}, 2021.

\bibitem[Seraj et~al.(2022)Seraj, Wang, Paleja, Martin, Sklar, Patel, and Gombolay]{seraj_2022_LearningEfficientDiversea}
E.~Seraj, Z.~Wang, R.~Paleja, D.~Martin, M.~Sklar, A.~Patel, and M.~Gombolay.
\newblock Learning efficient diverse communication for cooperative heterogeneous teaming.
\newblock In \emph{Proceedings of the 21st International Conference on Autonomous Agents and Multiagent Systems}, 2022.

\bibitem[Li et~al.(2021)Li, Wang, Wu, Zhao, Yang, and Zhang]{li_2021_CelebratingDiversityShared}
C.~Li, T.~Wang, C.~Wu, Q.~Zhao, J.~Yang, and C.~Zhang.
\newblock Celebrating diversity in shared multi-agent reinforcement learning.
\newblock In \emph{Advances in Neural Information Processing Systems}, 2021.

\bibitem[Wakilpoor et~al.(2020)Wakilpoor, Martin, Rebhuhn, and Vu]{wakilpoorHeterogeneousMultiAgentReinforcement2020}
C.~Wakilpoor, P.~J. Martin, C.~Rebhuhn, and A.~Vu.
\newblock Heterogeneous multi-agent reinforcement learning for unknown environment mapping.
\newblock \emph{arXiv preprint arXiv:2010.02663}, 2020.

\bibitem[Rashid et~al.(2020)Rashid, Samvelyan, de~Witt, Farquhar, Foerster, and Whiteson]{rashid2020monotonic}
T.~Rashid, M.~Samvelyan, C.~S. de~Witt, G.~Farquhar, J.~Foerster, and S.~Whiteson.
\newblock Monotonic value function factorisation for deep multi-agent reinforcement learning.
\newblock \emph{Journal of Machine Learning Research}, 2020.

\bibitem[ab~Abebe~Tessera et~al.(2025)ab~Abebe~Tessera, Rahman, and Albrecht]{tessera2025hypermarl}
K.~ab~Abebe~Tessera, A.~Rahman, and S.~V. Albrecht.
\newblock Hypermarl: Adaptive hypernetworks for multi-agent rl.
\newblock \emph{arXiv preprint arXiv:2412.04233}, 2025.

\bibitem[Oliehoek and Amato(2016)]{oliehoek2016concise}
F.~A. Oliehoek and C.~Amato.
\newblock \emph{A Concise Introduction to Decentralized POMDPs}.
\newblock Springer, 2016.

\bibitem[Ravichandar et~al.(2020)Ravichandar, Shaw, and Chernova]{ravichandar2020strata}
H.~Ravichandar, K.~Shaw, and S.~Chernova.
\newblock Strata: unified framework for task assignments in large teams of heterogeneous agents.
\newblock \emph{Autonomous Agents and Multi-Agent Systems}, 2020.

\bibitem[Yu et~al.(2022)Yu, Velu, Vinitsky, Gao, Wang, Bayen, and WU]{yuSurprisingEffectivenessPPO2022}
C.~Yu, A.~Velu, E.~Vinitsky, J.~Gao, Y.~Wang, A.~Bayen, and Y.~WU.
\newblock The surprising effectiveness of ppo in cooperative multi-agent games.
\newblock In \emph{Advances in Neural Information Processing Systems}, 2022.

\bibitem[Cho et~al.(2014)Cho, van Merrienboer, Gulcehre, Bahdanau, Bougares, Schwenk, and Bengio]{cho2014learningphraserepresentationsusing}
K.~Cho, B.~van Merrienboer, C.~Gulcehre, D.~Bahdanau, F.~Bougares, H.~Schwenk, and Y.~Bengio.
\newblock Learning phrase representations using rnn encoder-decoder for statistical machine translation.
\newblock \emph{arXiv preprint arXiv:1406.1078}, 2014.

\bibitem[Huang et~al.(2021)Huang, Xie, Bharadhwaj, and Shkurti]{huang_2021_ContinualModelBasedReinforcementa}
Y.~Huang, K.~Xie, H.~Bharadhwaj, and F.~Shkurti.
\newblock Continual model-based reinforcement learning with hypernetworks.
\newblock In \emph{2021 IEEE International Conference on Robotics and Automation (ICRA)}, 2021.

\bibitem[Rezaei-Shoshtari et~al.(2023)Rezaei-Shoshtari, Morissette, Hogan, Dudek, and Meger]{rezaei-shoshtari_2023_HypernetworksZeroshotTransfer}
S.~Rezaei-Shoshtari, C.~Morissette, F.~R. Hogan, G.~Dudek, and D.~Meger.
\newblock Hypernetworks for zero-shot transfer in reinforcement learning.
\newblock In \emph{Proceedings of the AAAI Conference on Artificial Intelligence}, 2023.

\bibitem[von Oswald et~al.(2020)von Oswald, Henning, Grewe, and Sacramento]{vonOswald_Henning_Grewe_Sacramento_2022}
J.~von Oswald, C.~Henning, B.~F. Grewe, and J.~Sacramento.
\newblock Continual learning with hypernetworks.
\newblock In \emph{International Conference on Learning Representations}, 2020.

\bibitem[Sarafian et~al.(2021)Sarafian, Keynan, and Kraus]{sarafian_2021_RecomposingReinforcementLearning}
E.~Sarafian, S.~Keynan, and S.~Kraus.
\newblock Recomposing the reinforcement learning building blocks with hypernetworks.
\newblock In \emph{Proceedings of the 38th International Conference on Machine Learning}, 2021.

\bibitem[Galanti and Wolf(2020)]{Galanti_Wolf_2020}
T.~Galanti and L.~Wolf.
\newblock On the modularity of hypernetworks.
\newblock In \emph{Advances in Neural Information Processing Systems}, 2020.

\bibitem[Beck et~al.(2023)Beck, Jackson, Vuorio, and Whiteson]{beck_vuorio_xiong_whiteson_2022}
J.~Beck, M.~T. Jackson, R.~Vuorio, and S.~Whiteson.
\newblock Hypernetworks in meta-reinforcement learning.
\newblock In \emph{Conference on Robot Learning}, 2023.

\bibitem[Chauhan et~al.(2024)Chauhan, Zhou, Lu, Molaei, and Clifton]{Chauhan_Zhou_Lu_Molaei_Clifton_2023}
V.~K. Chauhan, J.~Zhou, P.~Lu, S.~Molaei, and D.~A. Clifton.
\newblock A brief review of hypernetworks in deep learning.
\newblock \emph{Artificial Intelligence Review}, 2024.

\bibitem[Kraemer and Banerjee(2016)]{kraemer_2016_MultiagentReinforcementLearning}
L.~Kraemer and B.~Banerjee.
\newblock Multi-agent reinforcement learning as a rehearsal for decentralized planning.
\newblock \emph{Neurocomputing}, 2016.

\bibitem[Lowe et~al.(2017)Lowe, Wu, Tamar, Harb, Pieter~Abbeel, and Mordatch]{loweMultiAgentActorCriticMixed2017}
R.~Lowe, Y.~I. Wu, A.~Tamar, J.~Harb, O.~Pieter~Abbeel, and I.~Mordatch.
\newblock Multi-agent actor-critic for mixed cooperative-competitive environments.
\newblock \emph{Advances in Neural Information Processing Systems}, 2017.

\bibitem[Bettini et~al.(2024)Bettini, Shankar, and Prorok]{snd}
M.~Bettini, A.~Shankar, and A.~Prorok.
\newblock System neural diversity: Measuring behavioral heterogeneity in multi-agent learning.
\newblock \emph{arXiv preprint arXiv:2305.02128}, 2024.

\bibitem[Torbati et~al.(2023)Torbati, Lohiya, Singh, Nigam, and Ravichandar]{torbati2023marbler}
R.~J. Torbati, S.~Lohiya, S.~Singh, M.~S. Nigam, and H.~Ravichandar.
\newblock Marbler: An open platform for standardized evaluation of multi-robot reinforcement learning algorithms.
\newblock In \emph{2023 International Symposium on Multi-Robot and Multi-Agent Systems (MRS)}, 2023.

\bibitem[Papoudakis et~al.(2021)Papoudakis, Christianos, Schäfer, and Albrecht]{papoudakis2021benchmarking}
G.~Papoudakis, F.~Christianos, L.~Schäfer, and S.~V. Albrecht.
\newblock Benchmarking multi-agent deep reinforcement learning algorithms in cooperative tasks.
\newblock In \emph{Proceedings of the Neural Information Processing Systems Track on Datasets and Benchmarks (NeurIPS)}, 2021.

\bibitem[Ba et~al.(2016)Ba, Kiros, and Hinton]{ba2016layer}
J.~L. Ba, J.~R. Kiros, and G.~E. Hinton.
\newblock Layer normalization.
\newblock \emph{arXiv preprint arXiv:1607.06450}, 2016.

\bibitem[Ball et~al.(2023)Ball, Smith, Kostrikov, and Levine]{ball2023efficient}
P.~J. Ball, L.~Smith, I.~Kostrikov, and S.~Levine.
\newblock Efficient online reinforcement learning with offline data.
\newblock In \emph{International Conference on Machine Learning}, 2023.

\bibitem[Gallici et~al.(2025)Gallici, Fellows, Ellis, Pou, Masmitja, Foerster, and Martin]{gallici2024simplifying}
M.~Gallici, M.~Fellows, B.~Ellis, B.~Pou, I.~Masmitja, J.~N. Foerster, and M.~Martin.
\newblock Simplifying deep temporal difference learning.
\newblock In \emph{The Thirteenth International Conference on Learning Representations}, 2025.

\bibitem[Beck et~al.(2023)Beck, Vuorio, Xiong, and Whiteson]{Beck_neurips2023}
J.~Beck, R.~Vuorio, Z.~Xiong, and S.~Whiteson.
\newblock Recurrent hypernetworks are surprisingly strong in meta-rl.
\newblock In \emph{Advances in Neural Information Processing Systems}, 2023.

\bibitem[Kingma and Ba(2015)]{adam}
D.~Kingma and J.~Ba.
\newblock Adam: A method for stochastic optimization.
\newblock In \emph{International Conference on Learning Representations (ICLR)}, 2015.

\bibitem[Loshchilov and Hutter(2019)]{adamw}
I.~Loshchilov and F.~Hutter.
\newblock Decoupled weight decay regularization.
\newblock In \emph{International Conference on Learning Representations}, 2019.

\bibitem[Tieleman(2012)]{rmsprop}
T.~Tieleman.
\newblock Lecture 6.5-rmsprop: Divide the gradient by a running average of its recent magnitude.
\newblock \emph{COURSERA: Neural networks for machine learning}, 2012.

\end{thebibliography}

\clearpage
\appendix
\section{Evaluating Behavioral Diversity with SND}
\label{sec:SND}

To quantify the behavioral heterogeneity of the learned policies, we use the System Neural Diversity metric (SND) \cite{snd}.
While SND was introduced to evaluate the heterogeneity of independent policies trained without parameter sharing, we modify SND to evaluate the heterogeneity of parameter-shared policies conditioned on heterogeneous capabilities. Specifically, given an observation, we append each agent's capability to the observation and find the average pairwise distances between the policy outputs when conditioned on differing capabilities.
We define pairwise distance as
\begin{align*}
    d(i, j) = \frac{1}{|\mathcal{O}|} \sum_{o_t \in \mathcal{O}} TVD(\pi_{\theta}(o_t || c^i), \pi_{\theta}(o_t || c^j))
\end{align*}
where $\mathcal{O}$ is a set of observations obtained from several rollouts, $c^i$ is the capability vector of agent $i$, and $c^j$ is the capability vector of agent $j$. These pairwise distances are aggregated into a distance matrix, and the average of the upper triangular portion of this matrix is the final SND value. For value-based methods, individual networks output estimates of value per each action an agent can take. We interpret these Q-Values as a categorical distribution by taking a softmax over the values, and use Total Variational Distance when computing pairwise distances. Similarly, for discrete stochastic policies, we use Total Variational Distance between the categorical distributions over actions predicted by the policies.

\begin{figure}
    \begin{subfigure}{0.48\columnwidth}
      \includegraphics[width=\columnwidth]{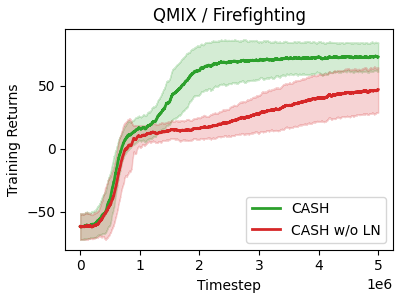}
    \end{subfigure}%
    \begin{subfigure}{0.48\columnwidth}
      \includegraphics[width=\columnwidth]{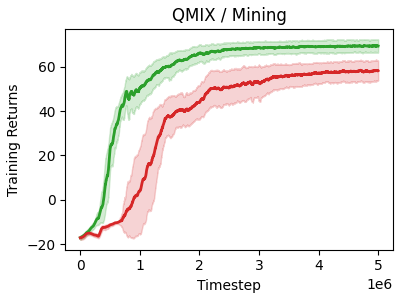}
    \end{subfigure}
    
    \begin{subfigure}{0.48\columnwidth}
      \includegraphics[width=\columnwidth]{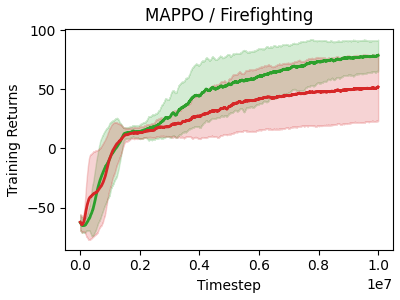}
    \end{subfigure}%
    \begin{subfigure}{0.48\columnwidth}
      \includegraphics[width=\columnwidth]{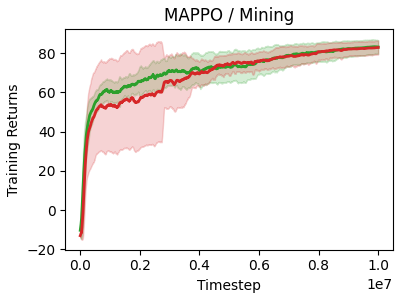}
    \end{subfigure}
    
    \begin{subfigure}{0.48\columnwidth}
      \includegraphics[width=\columnwidth]{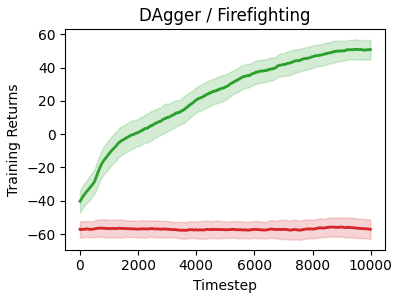}
    \end{subfigure}%
    \begin{subfigure}{0.48\columnwidth}
      \includegraphics[width=\columnwidth]{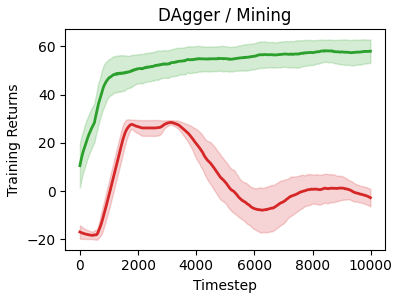}
    \end{subfigure}
    
    \caption{This figure shows the impact on training returns as a result of removing Layer Normalization from the Hyper Adapter of CASH. The results are presented for three learning paradigms across two simulation tasks. It is evident that LayerNorm is a crucial component in stabilizing the training of the hypernetwork within CASH.}
    \label{fig:ln-ablation}
\end{figure}

\section{Layer normalization is crucial to CASH}
\label{sec:layernorm}

To show the importance of adding layer normalization (LayerNorm) \cite{ba2016layer} when training hypernetworks for multi-agent coordination problems, we present an ablation over the inclusion of LayerNorm in CASH's Hyper Adapter in Figure \ref{fig:ln-ablation}.
Across all tasks and learning algorithms, the inclusion of LayerNorm improves training stability and decreases variance between seeds, which aligns with recent works suggesting the benefit of LayerNorm for improving training stability of deeper networks in RL~\cite{ball2023efficient, gallici2024simplifying}. 
In all cases except one (MAPPO / Mining), it results in a significant improvement on the final converged returns.
DAgger is a notable outlier in terms of performance gap with and without LayerNorm; we see that when training with DAgger, the performance of CASH without LayerNorm in \texttt{Mining} plummets around timestep 4000 and never recovers, while in \texttt{Firefighting}, CASH without LayerNorm never demonstrates better than random performance.

Taken together, these results show that layer normalization improves training stability for hypernetworks, and that this improved stability is crucial for our architecture. 
We note that adding layer normalization to hypernetworks was suggested by the original Hypernetworks paper \cite{Ha_Dai_Le_2016} to improve gradient flow, which our findings corroborate.
Surprisingly, prior work on using hypernetworks in meta-RL for generalization to new tasks \cite{beck_vuorio_xiong_whiteson_2022, Beck_neurips2023} does not include LayerNorm in the hypernetwork, even though we found it to be critically important for CASH.
We speculate that these works are able to achieve success despite omitting normalization from their hypernetworks because the input to their hypernetworks is a strong pretrained encoder.
By contrast, our hypernetwork is conditioned directly on task observations and team capabilities, without the benefit of an encoder for preprocessing.
We leave a more thorough investigation of normalization schemes for training deep hypernetworks to future work.

\section{Additional Training Details}
\label{sec:training-details}

\subsection{Hyper-Adapter}
\label{subsec:hyper-adapter}
Our experiments revealed that a deeper four-layer hypernetwork is better for our use case. 
However, consistent with prior work using hypernetworks in RL \cite{beck_vuorio_xiong_whiteson_2022}, we found it difficult to optimize a standard four-layer hypernetwork, both in reinforcement and imitation learning regimes.
We believe this is due to the combination of instabilities inherent in both the learning paradigms as well as hypernetworks themselves.
However, we found that simply adding layer normalization (LayerNorm)~\cite{ba2016layer} before each ReLU activation in our hypernetwork stabilizes learning and greatly improves performance (see Appendix \ref{sec:layernorm} for more details).

For CASH, we implemented the Hyper Adapter as two separate hypernetworks, one which generates the weights of the target linear layer and the other which generates the biases. We found that initializing both hypernetwork with orthogonal weights, zero-bias was conducive to good task success, which is common in other works. However, we noticed a slight performance increase when setting the \textit{scale} of those weights to be 0 for the bias-generating hypernet and 0.2 for weight-generating hypernet. We could not find a reference for this odd initialization scheme in the literature. 

\begin{table}
\centering
\begin{tabular}{|c|c|}
\hline
\textbf{Hyperparameter} & \textbf{Value} \\ \hline
Total timesteps & 10e6 \\ \hline
Learning rate & 2e-3 \\ \hline
Anneal learning rate & True \\ \hline
Update epochs & 4 \\ \hline
Number of minibatches & 4 \\ \hline
Gamma & 0.99 \\ \hline
GAE lambda & 0.95 \\ \hline
Clip epsilon & 0.2 \\ \hline
Scale clip epsilon & False \\ \hline
Entropy coefficient & 0.01 \\ \hline
Value function coefficient & 0.5 \\ \hline
Max gradient norm & 0.5 \\ \hline
Optimizer & Adam \cite{adam} \\ \hline
\end{tabular}
\caption{Hyperparameters for MAPPO}
\label{mappo-hyperparameters}
\end{table}

\begin{table}
\centering
\begin{tabular}{|c|c|}
\hline
\textbf{Hyperparameter} & \textbf{Value} \\ \hline
Total timesteps & 10e6 \\ \hline
Learning rate & 0.005 \\ \hline
Learning rate linear decay & True \\ \hline
Buffer size & 5000 \\ \hline
Buffer batch size & 32 \\ \hline
Epsilon start & 1.0 \\ \hline
Epsilon finish & 0.05 \\ \hline
Epsilon anneal time & 100000 \\ \hline
Mixer embedding dim & 32 \\ \hline
Mixer hypernetwork hidden dim & 64 \\ \hline
Mixer initialization scale & 0.00001 \\ \hline
Max gradient norm & 25 \\ \hline
Target update interval & 200 \\ \hline
Optimizer & AdamW \cite{adamw} \\ \hline
Epsilon Adam & 0.001 \\ \hline
Weight decay Adam & 0.00001 \\ \hline
TD lambda loss & True \\ \hline
TD lambda & 0.6 \\ \hline
Gamma & 0.9 \\ \hline
\end{tabular}
\caption{Hyperparameters for QMix (JaxMARL)}
\label{qmix-hyperparameters}
\end{table}

\begin{table}
\centering
\begin{tabular}{|c|c|}
\hline
\textbf{Hyperparameter} & \textbf{Value} \\ \hline
Total timesteps & 4e6 \\ \hline
Learning rate & 0.0005 \\ \hline
Learning rate linear decay & True \\ \hline
Buffer size & 5000 \\ \hline
Buffer batch size & 32 \\ \hline
Epsilon start & 1.0 \\ \hline
Epsilon finish & 0.05 \\ \hline
Epsilon anneal time & 50000 \\ \hline
Mixer embedding dim & 32 \\ \hline
Mixer hypernetwork hidden dim & 64 \\ \hline
Max gradient norm & 10 \\ \hline
Target update interval & 200 \\ \hline
Optimizer & RMSprop \cite{rmsprop} \\ \hline
Epsilon RMSProp & 0.00001 \\ \hline
Gamma & 0.9 \\ \hline
\end{tabular}
\caption{Hyperparameters for QMix (EPyMARL)}
\label{qmix-epm-hyperparameters}
\end{table}

\begin{table}
\centering
\begin{tabular}{|c|c|}
\hline
\textbf{Hyperparameter} & \textbf{Value} \\ \hline
Expert buffer size & 10000 \\ \hline
Initial expert trajectories & 1000 \\ \hline
Iterations & 10 \\ \hline
Trajectories per iteration & 1000 \\ \hline
Learning rate & 1e-4 \\ \hline
Learning rate linear decay & False \\ \hline
Beta & 1.0 \\ \hline
Beta linear decay & True \\ \hline
Updates per iteration & 100 \\ \hline
Update batch size & 64 \\ \hline
Max gradient norm & 1 \\ \hline
Optimizer & AdamW \cite{adamw} \\ \hline
Epsilon Adam & 0.001 \\ \hline
Weight decay Adam & 0 \\ \hline
\end{tabular}
\caption{Hyperparameters for DAgger}
\label{dagger-hyperparameters}
\end{table}

\subsection{MAPPO}

All networks are trained with the following hyperparameters in Table \ref{mappo-hyperparameters}. We ablated over four widths (32, 64, 128, 256) of RNN hidden dimension with the \texttt{RNN-IMP} architecture, and found 128 to be the best-performing variant as measured by test return. Based on this, we then designed a \texttt{CASH} architecture of lower parameter count than the baselines by selecting an RNN width of 32 and a hypernetwork width of 16.

\subsection{QMIX - JAXMARL}
All networks are trained with the following hyperparameters in Table \ref{qmix-hyperparameters}. We ablated over four widths (32, 64, 128, 256) of RNN hidden dimension with the \texttt{RNN-IMP} architecture, and found 128 to be the best-performing variant as measured by test return. Based on this, we then designed a \texttt{CASH} architecture of lower parameter count than the baselines by selecting an RNN width of 64 and a hypernetwork width of 32.

\subsection{QMIX - EPyMARL}
All networks are trained with the following hyperparameters in Table \ref{qmix-epm-hyperparameters}. We ablated over two widths (128, 256) of RNN hidden dimension with the \texttt{RNN-IMP} architecture, and found 256 to be the best-performing variant as measured by test return. Based on this, we then designed a \texttt{CASH} architecture of lower parameter count than the baselines by selecting a RNN width of 128 and a hypernetwork width of 64.

\subsection{DAgger}
All networks are trained with the following hyperparameters in Table \ref{dagger-hyperparameters}. We ablated over four widths (512, 1024, 2048, 4096) of RNN hidden dimension with the \texttt{RNN-IMP} architecture, and found 4096 to be the best-performing variant as measured by test return. We were unable to include larger width RNNs due to memory constraints. Based on this, we then designed a \texttt{CASH} architecture of lower parameter count than the baselines by selecting an RNN width of 2048 and a hypernetwork width of 1024.

\section{Capability Sampling Details}
\label{sec:cap-sample}
\subsection{Firefighting}
In  Table \ref{tab:fire-train-teams} and Table \ref{tab:fire-test-teams} we detail our randomly selected train and test teams for \texttt{Firefighting}. More details on how we chose these teams and how they were used can be found in the main body of Chapter 3. For context, in \texttt{Firefighting} the fires range from 0.2-0.3 strength and the training distribution of agents ranged between 0.1-0.3 firefighting capacity and 1-3 acceleration. For testing teams, we sample agents from out of bounds in both radius and acceleration, but match other agents to ensure task feasibility.

\begin{table}[h] 
    \centering
    \begin{tabular}{|c|c|c|}
        \hline
        Agent & (radius, acceleration) of agent \\
        \hline
        0 & (0.3, 1) \\
        1 & (0.2, 2) \\
        2 & (0.1, 3) \\
        3 & (0.1, 3) \\
        4 & (0.2, 2) \\
        \hline
    \end{tabular}
    \caption{\texttt{Firefighting} training robots.}
    \label{tab:fire-train-teams}
\end{table}

\begin{table}[h] 
    \centering
    \begin{tabular}{|c|c|}
        \hline
        Team & (radius, accel) for each agent in team \\
        \hline
        0 & (0.09, 3.43), (0.21, 2.94), (0.42, 0.75) \\
        1 & (0.09, 3.41), (0.21, 3.00), (0.48, 0.63) \\
        2 & (0.05, 3.46), (0.25, 2.76), (0.44, 0.60) \\
        3 & (0.08, 3.23), (0.23, 2.80), (0.50, 0.61) \\
        4 & (0.09, 3.14), (0.21, 1.16), (0.47, 0.86) \\
        5 & (0.06, 3.45), (0.24, 2.08), (0.46, 0.76) \\
        6 & (0.08, 3.06), (0.22, 1.08), (0.48, 0.56) \\
        7 & (0.07, 3.04), (0.23, 2.37), (0.45, 0.56) \\
        8 & (0.09, 3.36), (0.21, 2.20), (0.49, 0.64) \\
        9 & (0.09, 3.26), (0.21, 2.59), (0.47, 0.64) \\
        \hline
    \end{tabular}
    \caption{\texttt{Firefighting} testing teams.}
    \label{tab:fire-test-teams}
\end{table}

\subsection{Mining}
In \texttt{Mining} we have a training range between 0-0.5 for each capacity, with each agent having a total capacity across the two materials that sums to 0.5. During testing the random sampling is altered such that the total capacity of each agent sums to 1.0, to present an out of distribution challenge. Exact hyperparameters can be found in \ref{tab:Mining-train-teams} and \ref{tab:Mining-test-teams}.

\begin{table}[h]
    \centering
    \begin{tabular}{|c|c|}
        \hline
        Team & (1st material capacity, 2nd material capacity) for each agent of team \\
        \hline
        0 & (0.1, 0.4), (0.2, 0.3), (0.3, 0.2), (0.5, 0.0) \\
        1 & (0.2, 0.3), (0.3, 0.2), (0.4, 0.1), (0.5, 0.0) \\
        2 & (0.0, 0.5), (0.1, 0.4), (0.3, 0.2), (0.5, 0.0) \\
        3 & (0.0, 0.5), (0.1, 0.4), (0.2, 0.3), (0.4, 0.1) \\
        4 & (0.0, 0.5), (0.1, 0.4), (0.2, 0.3), (0.5, 0.0) \\
        5 & (0.1, 0.4), (0.3, 0.2), (0.4, 0.1), (0.5, 0.0) \\
        6 & (0.0, 0.5), (0.2, 0.3), (0.4, 0.1), (0.5, 0.0) \\
        7 & (0.0, 0.5), (0.2, 0.3), (0.3, 0.2), (0.5, 0.0) \\
        8 & (0.0, 0.5), (0.1, 0.4), (0.4, 0.1), (0.5, 0.0) \\
        9 & (0.1, 0.4), (0.2, 0.3), (0.4, 0.1), (0.5, 0.0) \\
        \hline
    \end{tabular}
    \caption{Mining training teams.}
    \label{tab:Mining-train-teams}
\end{table}

\begin{table}[h]
    \centering
    \begin{tabular}{|c|c|}
    \hline
    Team & (1st material capacity, 2nd material capacity) for each agent of team \\
    \hline
      0 & (0.72, 0.28), (0.98, 0.02), (0.17, 0.83), (0.12, 0.13) \\
      1 & (0.63, 0.37), (0.88, 0.12), (0.56, 0.44), (0.17, 0.08) \\
      2 & (0.04, 0.96), (0.24, 0.76), (0.29, 0.71), (0.25, 0.  ) \\
      3 & (0.01, 0.99), (0.56, 0.44), (0.18, 0.82), (0.05, 0.2 ) \\
      4 & (0.26, 0.74), (0.65, 0.35), (0.95, 0.05), (0.16, 0.09) \\
      5 & (0.37, 0.63), (0.52, 0.48), (0.77, 0.23), (0.04, 0.21) \\
      6 & (0.41, 0.59), (0.96, 0.04), (0.12, 0.88), (0.05, 0.2 ) \\
      7 & (0.86, 0.14), (0.97, 0.03), (0.38, 0.62), (0.12, 0.13) \\
      8 & (0.68, 0.32), (0.87, 0.13), (0.73, 0.27), (0.15, 0.1 ) \\
      9 & (0.4,  0.6),  (0.41, 0.59), (0.54, 0.46), (0.02, 0.23) \\
    
    \hline
    \end{tabular}
    \caption{Mining testing teams.}
    \label{tab:Mining-test-teams}
\end{table}

\subsection{Material Transport}
In \texttt{Material Transport} we construct teams of 4 agents consisting of,
\begin{itemize}
    \item 2 \textit{fast} agents with \textit{low} capacity.
    \item 2 \textit{slow} with \textit{high} capacity.
\end{itemize}
The values for each are uniformly sampled from the sets in Table \ref{tab:MT-teams}. We deploy the trained policies on the five teams in Table \ref{tab:robotarium-MT}, where the values for each capability were randomly sampled from the training capability sets.
\begin{table}[h]
    \centering
    \begin{tabular}{|c|c|}
    \hline
        \textit{fast} & 0.40, 0.45, 0.50, 0.55, 0.60 \\
        \textit{slow} & 0.10, 0.15, 0.20, 0.25, 0.30 \\
        \textit{high} & 14, 16, 18, 20, 22 \\
        \textit{low} & 4, 6, 8, 10, 12 \\
    \hline
    \end{tabular}
    \caption{Material transport capability sets.}
    \label{tab:MT-teams}
\end{table}

\begin{table}[h]
    \centering
    \begin{tabular}{|c|c|}
    \hline
    Team & (speed, capacity) for each robot \\
    \hline
      0 & (0.25, 20), (0.25, 20), (0.4, 6), (0.4, 6) \\
      1 & (0.30, 22), (0.30, 22), (0.55, 6), (0.55, 6) \\
      2 & (0.20, 20), (0.20, 20), (0.55, 6), (0.55, 6) \\
      3 & (0.30, 16), (0.30, 16), (0.50, 12), (0.50, 12) \\
      4 & (0.10, 14), (0.10, 14), (0.60, 12), (0.60, 12) \\
    
    \hline
    \end{tabular}
    \caption{Material transport deployment teams.}
    \label{tab:robotarium-MT}
\end{table}

\subsection{Predator Capture Prey}
In \texttt{Predator Capture Prey} we construct teams of 4 agents consisting of 2 \texttt{capture} agents and 2 \texttt{sensing} agents. The capabilities for each are uniformly sampled from the sets in Table \ref{tab:PCP-teams}. We deploy the trained policies on the five teams in Table \ref{tab:robotarium-PCP}, where the values for each capability were randomly sampled from the training capability sets.
\begin{table}[h]
    \centering
    \begin{tabular}{|c|c|}
    \hline
        \textit{capture radii} & 0.10, 0.125, 0.15, 0.175, .20 \\
        \textit{sensing radii} & 0.20, 0.25, 0.30, 0.35, 0.40 \\
    \hline
    \end{tabular}
    \caption{Predator capture prey capability sets.}
    \label{tab:PCP-teams}
\end{table}

\begin{table}[h]
    \centering
    \begin{tabular}{|c|c|}
    \hline
    Team & (capture radius, sensing radius) for each robot \\
    \hline
      0 & (0.15, 0.00), (0.15, 0.00), (0.00, 0.20), (0.00, 0.20) \\
      1 & (0.15, 0.00), (0.15, 0.00), (0.00, 0.35), (0.00, 0.35) \\
      2 & (0.125, 0.00), (0.125, 0.00), (0.00, 0.30), (0.00, 0.30) \\
      3 & (0.20, 0.00), (0.20, 0.00), (0.00, 0.40), (0.00, 0.40) \\
      4 & (0.10, 0.00), (0.10, 0.00), (0.00, 0.35), (0.00, 0.35) \\
    
    \hline
    \end{tabular}
    \caption{Predator capture prey deployment teams.}
    \label{tab:robotarium-PCP}
\end{table}

\section{Environment Implementations}
\label{sec:env-implementations}

We extended JaxMARL to include our two custom environments as well as a training script which implements DAgger. We use the implemented MARBLER Material Transport and Predator Capture Prey tasks, and extend them to sample from capability sets.

Please refer to our code for additional information on reward schemes, observation spaces, etc.

\section{Additional MARBLER Results}
\begin{figure}[h]
    \centering
    \begin{subfigure}{0.33\columnwidth}
      \includegraphics[width=\columnwidth]{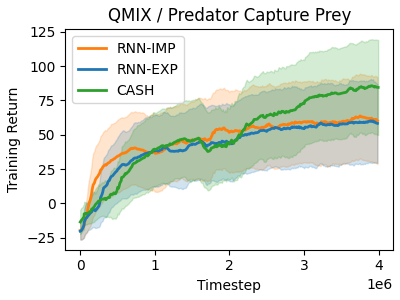}
    \end{subfigure}
    \begin{subfigure}{0.33\columnwidth}
      \includegraphics[width=\columnwidth]{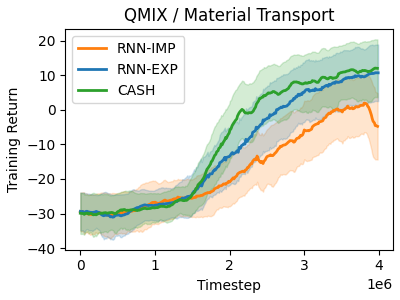}
    \end{subfigure}
    
    \caption{MARBLER training curves for \texttt{Predator Capture Prey} and \texttt{Material Transport}.}
    \label{fig:marbler-returns}
\end{figure}

We report the training curves for \texttt{Material Transport} and \texttt{Predator Capture Prey} in Fig. \ref{fig:marbler-returns}. The plots are smoothed by downsampling the original mean and standard deviation over all seeds and then taking a rolling average.

\subsection{Team Size Generalization}
\label{sec:marbler-12-agent}

We leverage the decentralized execution of CASH's policy design to zero-shot deploy the policies trained on 4-robot teams on teams of 12 robots. We scale both environments by doubling the amount of material or the number of prey, and robots are restricted to observing their three nearest neighbors.

\begin{table*}[h] 
    \centering
    \resizebox{=0.7\textwidth}{!}{
    \begin{tabular}{|c|c|c|c|}
        \hline
        & &\texttt{MT} &\texttt{PCP} \\
        Scenario & Architecture & Task Completion Rate ($\uparrow$) & Task Completion Rate ($\uparrow$) \\
        \hline
        \texttt{Zero-Shot} & RNN-IMP & 0.45 $\pm$ 0.39 & 0.67 $\pm$ 0.21 \\
        (4 $\rightarrow$ 12)& RNN-EXP & 0.32 $\pm$ 0.39  & 0.58 $\pm$ 0.25 \\
        & CASH & \textbf{0.67 $\pm$ 0.40} & \textbf{0.68 $\pm$ 0.21} \\
        \hline
    \end{tabular}
    }
    \caption{
    Zero-shot decentralized deployment to 12 robots over 50 episodes across 3 random seeds. 
    Task completion is the percentage of prey captured or material dropped off.
    }
    \label{tab:marbler-scale}
\end{table*}

\textbf{CASH can be readily decentralized and deployed on larger robot teams.}
As seen in Table \ref{tab:marbler-scale}, CASH policies can be zero-shot deployed on larger teams. CASH performs either similarly to (in \texttt{PCP}) or better than (in \texttt{MT}) the baselines in when deployed on teams of 12 robots. The modest improvement over RNN-IMP in \texttt{PCP} be explained by the fact that increasing the density of predators and prey decreases the need for tight coordination among the predators.

\section{Additional JaxMARL Results}
\label{sec:additional-jaxmarl-results}
\begin{figure}[h]
    \begin{subfigure}{0.33\columnwidth}
      \includegraphics[width=\columnwidth]{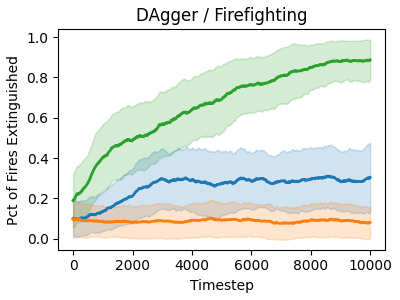}
    \end{subfigure}
    \begin{subfigure}{0.33\columnwidth}
      \includegraphics[width=\columnwidth]{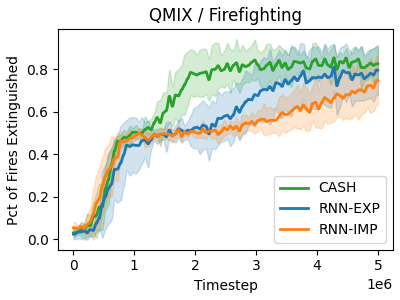}
    \end{subfigure}%
    \begin{subfigure}{0.33\columnwidth}
      \includegraphics[width=\columnwidth]{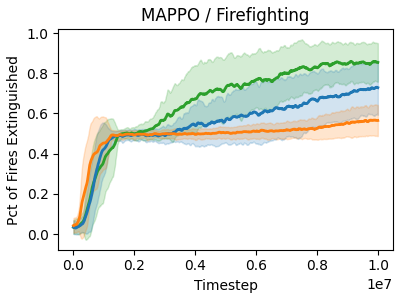}
    \end{subfigure}%
    
    \begin{subfigure}{0.33\columnwidth}
      \includegraphics[width=\columnwidth]{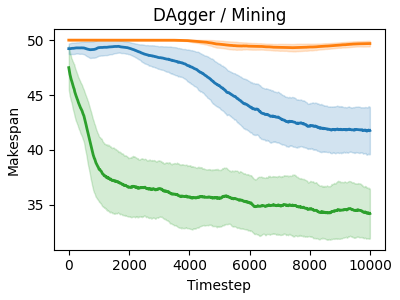}
    \end{subfigure}%
    \begin{subfigure}{0.33\columnwidth}
      \includegraphics[width=\columnwidth]{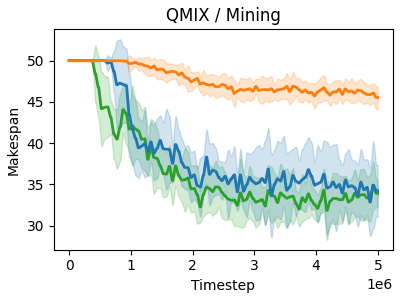}
    \end{subfigure}%
    \begin{subfigure}{0.33\columnwidth}
      \includegraphics[width=\columnwidth]{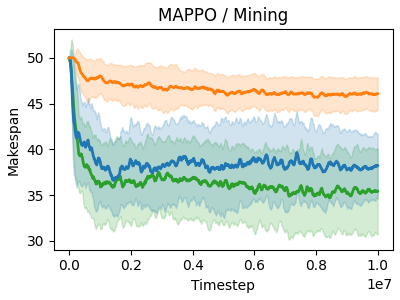}
    \end{subfigure}
    
    \caption{Task performance metrics across two JaxMARL tasks and three learning paradigms as evaluated on out-of-distribution robot capabilities and unseen team compositions. Percentage of Fires Extinguished ($\uparrow$, top row) is for Firefighting. Makespan ($\downarrow$, bottom row) is for Mining. These metrics provide additional context beyond the success rates in Table \ref{tab:success_rates}.}
    \label{fig:task-metrics}
\end{figure}

\begin{figure}[h]
    \begin{subfigure}{0.33 \columnwidth}
      \includegraphics[width=\columnwidth]{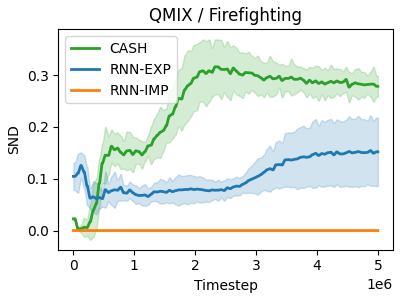}
    \end{subfigure}%
    \begin{subfigure}{0.33 \columnwidth}
      \includegraphics[width=\columnwidth]{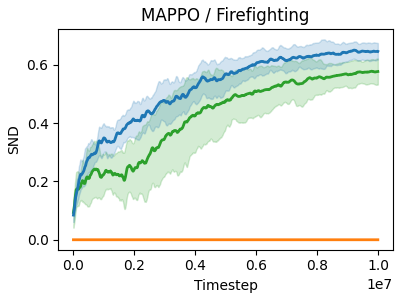}
    \end{subfigure}%
    \begin{subfigure}{0.33\columnwidth}
      \includegraphics[width=\columnwidth]{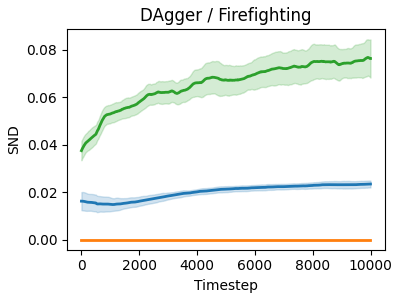}
    \end{subfigure}

    \begin{subfigure}{0.33\columnwidth}
      \includegraphics[width=\columnwidth]{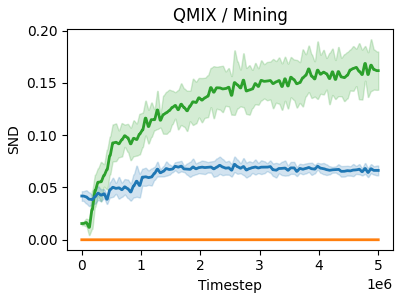}
    \end{subfigure}%
    \begin{subfigure}{0.33\columnwidth}
      \includegraphics[width=\columnwidth]{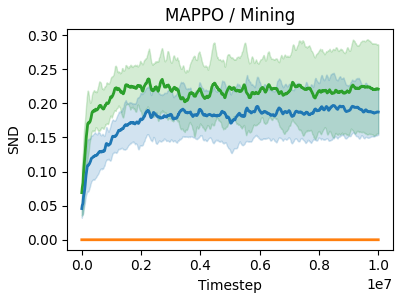}
    \end{subfigure} %
    \begin{subfigure}{0.33\columnwidth}
      \includegraphics[width=\columnwidth]{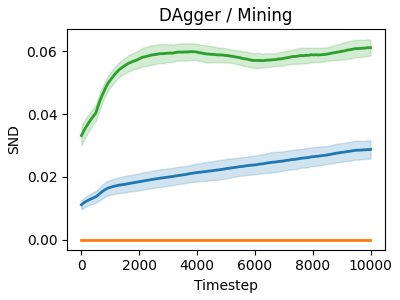}
    \end{subfigure}
    
    \caption{
    Behavioral diversity ($\updownarrow$) across two JaxMARL tasks and three learning paradigms.
    In all cases but one (MAPPO/\texttt{Firefighting}), CASH learns more diverse behaviors than the baseline approaches. 
    Qualitative analysis of MAPPO/\texttt{Firefighting} suggests that RNN-EXP is able to learn diverse behaviors, but the corresponding success rate in Table~\ref{tab:success_rates} suggests these behaviors are unproductive for task success. Conversely, CASH is able to learn an appropriate level of heterogeneity for strong task success.}
    \label{fig:snd}
\end{figure}

\subsection{Comparison against ID-based methods}
\label{sec:jaxmarl-id}
\begin{figure}
    \begin{subfigure}{0.24 \columnwidth}
      \includegraphics[width=\columnwidth]{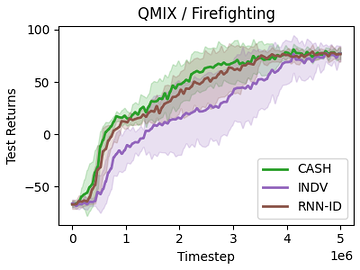}
    \end{subfigure}%
    \begin{subfigure}{0.24 \columnwidth}
      \includegraphics[width=\columnwidth]{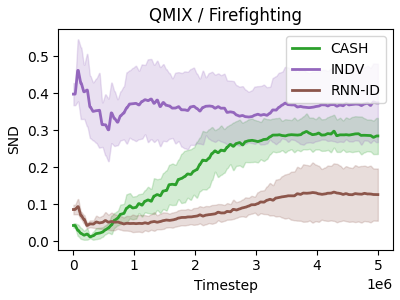}
    \end{subfigure}%
    \begin{subfigure}{0.24\columnwidth}
      \includegraphics[width=\columnwidth]{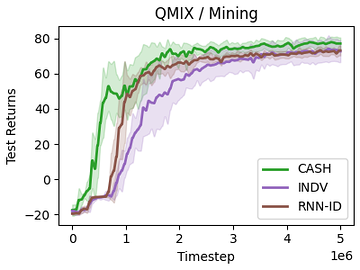}
    \end{subfigure}
    \begin{subfigure}{0.24\columnwidth}
      \includegraphics[width=\columnwidth]{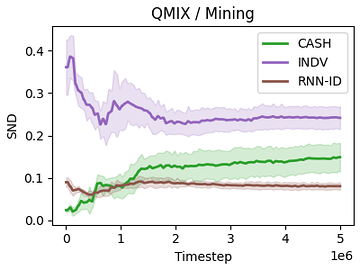}
    \end{subfigure}
    \caption{ID-based methods are more sample efficient than individualized policies (see returns), but learn a lower level of behavioral diversity (see SND). CASH matches or outperforms ID-based methods in both sample efficiency and behavioral diversity.}
    \label{fig:id-ablation}
\end{figure}

To further contextualize the \texttt{INDV} results, we compare against shared-parameter ID-based methods where a unique one-hot ID is appended to the observation space of each agent. ID-based methods represent a shared parameter baseline that cannot generalize to unseen robots. The experiment setup is identical to that in Section \ref{subsec:cash-vs-indp}, but includes results from an agent ID baseline labeled \texttt{RNN-ID}. We plot the returns and SND for the \texttt{Firefighting} and \texttt{Mining} tasks in Figure \ref{fig:id-ablation}. Expectedly, the shared encoder does improve sample efficiency when compared to individualized policies, and \texttt{RNN-ID} learns a lower level of behavioral diversity.

\subsection{Training on Larger Teams}
\label{sec:jaxmarl-12-agent}
We investigate training CASH, RNN-EXP, and RNN-IMP on 12 agent teams in \texttt{Mining} and \texttt{Firefighting}. We restrict robots in both scenarios to only observe their three nearest neighbors to model decentralized execution.

\subsubsection{Scaled Environments}
For \texttt{Firefighting}, we increased the strength of fires to range from 0.3-0.4, and the distribution of agents ranged between 0.05-0.2 firefighting capacity and 1-3 acceleration. We sample testing teams as described in Appendix \ref{sec:cap-sample}. Exact training and testing teams are listed in Tables \ref{tab:fire-scale-train-teams} and \ref{tab:fire-scale-test-teams}.

For \texttt{Mining}, the material quotas scale with the number of agents; see our provided code for exact implementations. We retain the same sampling strategy for constructing training (in-distribution) and testing (out-of-distribution) teams as described in Appendix \ref{sec:cap-sample}. Exact teams are listed in Tables \ref{tab:mining-scale-train-teams} and \ref{tab:mining-scale-test-teams}.

\subsubsection{Training Details}
We train all baselines with QMIX and MAPPO in both environments. We choose the hidden width for RNN-EXP and RNN-IMP by ablating over widths and learning rates for RNN-IMP and selecting the best performing hidden width and learning rate combination. For QMIX, we use hidden widths (64, 128, 256) and learning rates (0.005,0.003,0.001). For MAPPO, we use hidden widths (64, 128, 256) and learning rates (0.002, 0.001, 0.0005, 0.0003). We then construct CASH using a smaller hidden width and perform the same learning rate ablation. We report the exact widths and learning rates of each baseline for each algorithm and environment in Table \ref{tab:scale-hyperparams}. For CASH, we report the (hidden width, hypernetwork hidden width) respectively.

\begin{table*}[h]
    \centering
    \begin{tabular}{|c|c|c|c|c|}
        \hline
        Environment & Algorithm & Architecture & Width & Learning Rate \\
        \hline
        \texttt{Firefighting} & QMIX    & CASH     & (64, 32) & 0.003 \\
        (12-agent)            &         & RNN-EXP  & 128  & 0.003 \\
                              &         & RNN-IMP  & 128  & 0.003 \\
                              & MAPPO   & CASH     & (64, 32) & 0.0003 \\
                              &         & RNN-EXP  & 256  & 0.001 \\
                              &         & RNN-IMP  & 256  & 0.001 \\
        \hline
        \texttt{Mining} & QMIX    & CASH     & (128, 64) & 0.003 \\
        (12-agent)            &         & RNN-EXP  & 256  & 0.003 \\
                              &         & RNN-IMP  & 256  & 0.003 \\
                              & MAPPO   & CASH     & (64, 32) & 0.0005 \\
                              &         & RNN-EXP  & 256  & 0.0005 \\
                              &         & RNN-IMP  & 256  & 0.0005 \\
        \hline
    \end{tabular}
    \caption{Final hidden widths and learning rates used for 12 agent \texttt{Firefighting} and \texttt{Mining}.}
    \label{tab:scale-hyperparams}
\end{table*}

\subsubsection{Results}
\begin{table*}[h]
    \centering
    \resizebox{\textwidth}{!}{%
    \begin{tabular}{|c|c|c|ccc|}
        \hline
        Algorithm / & Architecture & \# Parameters & In-Distribution & Out-of-Distribution & Out-of-Distribution \\
        Environment & &($\downarrow$) &Success Rate ($\uparrow$) & Success Rate ($\uparrow$) & SND \\
        \hline
        QMIX /                  & RNN-IMP & 103K & 0.17 $\pm$ 0.30 & 0.08 $\pm$ 0.08 & 0.00 $\pm$ 0.00 \\
        \texttt{Firefighting}   & RNN-EXP & 103K & 0.50 $\pm$ 0.58 & 0.09 $\pm$ 0.11 & 0.08 $\pm$ 0.04 \\
        (12-agent)              & CASH & \textbf{43K} & \textbf{0.97 $\pm$ 0.02} & \textbf{0.33 $\pm$ 0.19} &0.17 $\pm$ 0.03 \\
        \hline
        MAPPO /                 & RNN-IMP & 468K & 0.15 $\pm$ 0.10 & 0.31 $\pm$ 0.06 & 0.00 $\pm$ 0.00 \\
        \texttt{Firefighting}   & RNN-EXP & 468K & 0.75 $\pm$ 0.23 & 0.29 $\pm$ 0.04 & 0.31 $\pm$ 0.02 \\
        (12-agent)              & CASH & \textbf{170K} & \textbf{1.00 $\pm$ 0.00} & \textbf{0.54 $\pm$ 0.10} &0.36 $\pm$ 0.02 \\
        \hline
        QMIX /                  & RNN-IMP & 403K & 0.91 $\pm$ 0.05 & 0.89 $\pm$ 0.11 & 0.00 $\pm$ 0.00 \\
        \texttt{Mining}         & RNN-EXP & 403K & 0.93 $\pm$ 0.07 & 0.87 $\pm$ 0.22 & 0.08 $\pm$ 0.01 \\
        (12-agent)              & CASH & \textbf{165K} & \textbf{0.94 $\pm$ 0.05} & \textbf{0.94 $\pm$ 0.10} &0.17 $\pm$ 0.01 \\
        \hline
        MAPPO /                 & RNN-IMP & 469K & 0.81 $\pm$ 0.06 & \textbf{0.90 $\pm$ 0.10} & 0.00 $\pm$ 0.00 \\
        \texttt{Mining}         & RNN-EXP & 469K & 0.96 $\pm$ 0.07 & 0.88 $\pm$ 0.11 & 0.09 $\pm$ 0.03 \\
        (12-agent)              & CASH & \textbf{170K} & \textbf{0.98 $\pm$ 0.04} & 0.71 $\pm$ 0.34 & 0.12 $\pm$ 0.03 \\
        \hline
    \end{tabular}}
    \caption{
    CASH tends to achieve the highest success rates across learning paradigms on the 12-agent JaxMARL tasks. CASH also seems to retain the generalization benefits observed in Table \ref{tab:success_rates}, though we acknowledge MAPPO / \texttt{Mining} is a notable exception. 
    }
    \label{tab:jaxmarl-scale}
\end{table*}

We find that CASH consistently outperforms baselines in-distribution when training on large teams. Consistent with section \ref{sec:jaxmarl_experiments}, we find that CASH tends to exhibit more diverse behaviors as well. When generalizing to out-of-distribution teams, we find that CASH outperforms baselines across all conditions except on MAPPO/\texttt{Mining}, where surprisingly RNN-IMP. Future work can investigate under what conditions CASH is less effective in learning for larger team sizes. Likely, curriculum approaches commonly employed in multi-agent scenarios could be useful in learning more robust strategies with large teams.

\begin{table*}[h]
    \centering
    \begin{tabular}{|c|c|}
        \hline
        Agent & (radius, acceleration) of agent \\
        \hline
        0 & (0.07, 2.31) \\
        1 & (0.09, 1.74) \\
        2 & (0.07, 2.63) \\
        3 & (0.06, 2.84) \\
        4 & (0.06, 1.33) \\
        5 & (0.10, 2.66) \\
        6 & (0.09, 1.89) \\
        7 & (0.09, 2.98) \\
        8 & (0.15, 2.48) \\
        9 & (0.19, 2.13) \\
        10 & (0.19, 2.56) \\
        11 & (0.16, 1.99) \\
        12 & (0.07, 2.31) \\
        13 & (0.09, 1.74) \\
        14 & (0.07, 2.63) \\
        15 & (0.06, 2.84) \\
        16 & (0.06, 1.33) \\
        17 & (0.10, 2.66) \\
        18 & (0.09, 1.89) \\
        19 & (0.09, 2.98) \\
        20 & (0.15, 2.48) \\
        21 & (0.19, 2.13) \\
        22 & (0.19, 2.56) \\
        23 & (0.16, 1.99) \\
        \hline
    \end{tabular}
    \caption{12 agent \texttt{Firefighting} training teams.}
    \label{tab:fire-scale-train-teams}
\end{table*}

\begin{table*}[h]
    \centering
    \resizebox{\textwidth}{!}{%
    \begin{tabular}{|c|c|}
        \hline
        Team & (radius, accel) for each agent in team \\
        \hline
        0 & (0.04, 3.49), (0.04, 3.34), (0.04, 3.33), (0.04, 3.01), (0.10, 2.96), (0.11, 1.28), (0.10, 2.97), (0.10, 2.35), (0.23, 0.62), (0.23, 0.81), (0.22, 0.52), (0.25, 0.76) \\
        1 & (0.05, 3.00), (0.04, 3.23), (0.03, 3.43), (0.05, 3.35), (0.10, 1.51), (0.11, 1.89), (0.11, 1.56), (0.10, 1.97), (0.22, 0.92), (0.24, 0.97), (0.20, 0.82), (0.22, 0.64) \\
        2 & (0.03, 3.04), (0.03, 3.15), (0.04, 3.35), (0.04, 3.03), (0.11, 2.50), (0.11, 2.71), (0.10, 1.78), (0.11, 2.63), (0.20, 0.52), (0.22, 0.74), (0.21, 0.89), (0.22, 0.57) \\
        3 & (0.05, 3.25), (0.05, 3.47), (0.04, 3.23), (0.04, 3.02), (0.10, 2.23), (0.10, 2.06), (0.10, 2.19), (0.10, 2.61), (0.21, 0.81), (0.22, 0.76), (0.24, 0.51), (0.24, 0.73) \\
        4 & (0.03, 3.06), (0.04, 3.47), (0.03, 3.44), (0.04, 3.45), (0.11, 1.82), (0.11, 2.27), (0.11, 2.76), (0.10, 2.38), (0.21, 0.90), (0.21, 0.64), (0.21, 0.90), (0.22, 0.93) \\
        5 & (0.04, 3.11), (0.04, 3.11), (0.04, 3.20), (0.03, 3.33), (0.11, 1.94), (0.10, 1.06), (0.11, 2.10), (0.11, 1.17), (0.21, 0.58), (0.24, 0.93), (0.20, 0.59), (0.21, 0.52) \\
        6 & (0.03, 3.35), (0.03, 3.38), (0.04, 3.14), (0.04, 3.22), (0.11, 1.05), (0.11, 1.35), (0.10, 2.15), (0.11, 1.59), (0.23, 0.80), (0.21, 0.94), (0.20, 0.71), (0.21, 0.56) \\
        7 & (0.03, 3.14), (0.03, 3.43), (0.03, 3.34), (0.03, 3.31), (0.11, 1.07), (0.11, 1.21), (0.11, 2.74), (0.11, 1.14), (0.21, 0.96), (0.24, 0.98), (0.23, 0.69), (0.23, 0.74) \\
        8 & (0.04, 3.02), (0.04, 3.05), (0.05, 3.16), (0.03, 3.14), (0.10, 2.02), (0.10, 1.78), (0.10, 1.24), (0.11, 1.48), (0.21, 0.63), (0.20, 0.85), (0.21, 0.64), (0.22, 0.54) \\
        9 & (0.04, 3.32), (0.04, 3.16), (0.04, 3.07), (0.03, 3.11), (0.11, 1.66), (0.10, 1.98), (0.10, 2.79), (0.11, 2.61), (0.24, 0.84), (0.21, 0.78), (0.22, 0.76), (0.21, 0.55) \\
        \hline
    \end{tabular}}
    \caption{12 agent \texttt{Firefighting} testing teams.}
    \label{tab:fire-scale-test-teams}
\end{table*}

\begin{table*}[h]
    \centering
    \resizebox{\textwidth}{!}{%
    \begin{tabular}{|c|c|}
        \hline
        Team & (1st material capacity, 2nd material capacity) for each agent of team \\
        \hline
        0 & (0.00, 0.50), (0.06, 0.44), (0.09, 0.41), (0.18, 0.32), (0.21, 0.29), (0.24, 0.26), (0.26, 0.24), (0.29, 0.21), (0.32, 0.18), (0.35, 0.15), (0.38, 0.12), (0.41, 0.09) \\
        1 & (0.00, 0.50), (0.03, 0.47), (0.06, 0.44), (0.09, 0.41), (0.12, 0.38), (0.21, 0.29), (0.29, 0.21), (0.32, 0.18), (0.35, 0.15), (0.44, 0.06), (0.47, 0.03), (0.50, 0.00) \\
        2 & (0.00, 0.50), (0.03, 0.47), (0.12, 0.38), (0.18, 0.32), (0.26, 0.24), (0.29, 0.21), (0.32, 0.18), (0.35, 0.15), (0.38, 0.12), (0.41, 0.09), (0.44, 0.06), (0.50, 0.00) \\
        3 & (0.00, 0.50), (0.09, 0.41), (0.15, 0.35), (0.18, 0.32), (0.21, 0.29), (0.24, 0.26), (0.26, 0.24), (0.29, 0.21), (0.41, 0.09), (0.44, 0.06), (0.47, 0.03), (0.50, 0.00) \\
        4 & (0.00, 0.50), (0.09, 0.41), (0.12, 0.38), (0.15, 0.35), (0.18, 0.32), (0.24, 0.26), (0.32, 0.18), (0.38, 0.12), (0.41, 0.09), (0.44, 0.06), (0.47, 0.03), (0.50, 0.00) \\
        5 & (0.06, 0.44), (0.09, 0.41), (0.12, 0.38), (0.15, 0.35), (0.18, 0.32), (0.24, 0.26), (0.26, 0.24), (0.32, 0.18), (0.35, 0.15), (0.38, 0.12), (0.44, 0.06), (0.47, 0.03) \\
        6 & (0.00, 0.50), (0.03, 0.47), (0.06, 0.44), (0.15, 0.35), (0.18, 0.32), (0.21, 0.29), (0.26, 0.24), (0.29, 0.21), (0.35, 0.15), (0.41, 0.09), (0.47, 0.03), (0.50, 0.00) \\
        7 & (0.06, 0.44), (0.09, 0.41), (0.12, 0.38), (0.15, 0.35), (0.21, 0.29), (0.26, 0.24), (0.29, 0.21), (0.35, 0.15), (0.38, 0.12), (0.41, 0.09), (0.47, 0.03), (0.50, 0.00) \\
        8 & (0.00, 0.50), (0.03, 0.47), (0.09, 0.41), (0.12, 0.38), (0.18, 0.32), (0.24, 0.26), (0.26, 0.24), (0.35, 0.15), (0.41, 0.09), (0.44, 0.06), (0.47, 0.03), (0.50, 0.00) \\
        9 & (0.00, 0.50), (0.03, 0.47), (0.06, 0.44), (0.09, 0.41), (0.18, 0.32), (0.21, 0.29), (0.26, 0.24), (0.29, 0.21), (0.32, 0.18), (0.35, 0.15), (0.41, 0.09), (0.47, 0.03) \\
        \hline
    \end{tabular}}
    \caption{12 agent \texttt{Mining} training teams.}
    \label{tab:mining-scale-train-teams}
\end{table*}

\begin{table*}[h]
    \centering
    \resizebox{\textwidth}{!}{%
    \begin{tabular}{|c|c|}
        \hline
        Team & (1st material capacity, 2nd material capacity) for each agent of team \\
        \hline
        0 & (0.31, 0.69), (0.75, 0.25), (0.47, 0.53), (0.70, 0.30), (0.61, 0.39), (0.62, 0.38), (0.44, 0.56), (0.43, 0.57), (0.99, 0.01), (0.11, 0.14), (0.10, 0.15), (0.10, 0.15) \\
        1 & (0.34, 0.66), (0.40, 0.60), (0.93, 0.07), (0.91, 0.09), (0.77, 0.23), (0.31, 0.69), (0.11, 0.89), (0.73, 0.27), (0.79, 0.21), (0.04, 0.21), (0.06, 0.19), (0.12, 0.13) \\
        2 & (0.74, 0.26), (0.51, 0.49), (0.48, 0.52), (0.92, 0.08), (0.06, 0.94), (0.84, 0.16), (0.37, 0.63), (0.87, 0.13), (0.47, 0.53), (0.16, 0.09), (0.02, 0.23), (0.03, 0.22) \\
        3 & (0.32, 0.68), (0.85, 0.15), (0.27, 0.73), (0.43, 0.57), (0.23, 0.77), (0.49, 0.51), (0.62, 0.38), (0.57, 0.43), (0.88, 0.12), (0.23, 0.02), (0.23, 0.02), (0.05, 0.20) \\
        4 & (0.94, 0.06), (0.98, 0.02), (0.39, 0.61), (0.30, 0.70), (0.96, 0.04), (0.82, 0.18), (0.35, 0.65), (0.90, 0.10), (0.56, 0.44), (0.13, 0.12), (0.16, 0.09), (0.01, 0.24) \\
        5 & (0.48, 0.52), (0.70, 0.30), (0.91, 0.09), (0.57, 0.43), (0.87, 0.13), (0.44, 0.56), (0.09, 0.91), (0.84, 0.16), (0.58, 0.42), (0.16, 0.09), (0.19, 0.06), (0.04, 0.21) \\
        6 & (0.08, 0.92), (0.66, 0.34), (0.73, 0.27), (0.67, 0.33), (0.33, 0.67), (0.69, 0.31), (0.09, 0.91), (0.18, 0.82), (0.79, 0.21), (0.09, 0.16), (0.12, 0.13), (0.05, 0.20) \\
        7 & (0.45, 0.55), (0.20, 0.80), (0.93, 0.07), (0.32, 0.68), (0.63, 0.37), (0.59, 0.41), (0.82, 0.18), (0.56, 0.44), (0.73, 0.27), (0.01, 0.24), (0.02, 0.23), (0.20, 0.05) \\
        8 & (0.16, 0.84), (0.55, 0.45), (0.12, 0.88), (0.54, 0.46), (0.70, 0.30), (0.40, 0.60), (0.57, 0.43), (1.00, 0.00), (0.52, 0.48), (0.05, 0.20), (0.06, 0.19), (0.01, 0.24) \\
        9 & (0.87, 0.13), (0.14, 0.86), (0.68, 0.32), (0.19, 0.81), (0.52, 0.48), (0.85, 0.15), (0.73, 0.27), (0.88, 0.12), (0.60, 0.40), (0.05, 0.20), (0.05, 0.20), (0.13, 0.12) \\
        \hline
    \end{tabular}}
    \caption{12 agent \texttt{Mining} testing teams.}
    \label{tab:mining-scale-test-teams}
\end{table*}

\end{document}